\pdfoutput=1

\documentclass[journal,onecolumn]{IEEEtran}

\usepackage[utf8x]{inputenc}
\usepackage{amsmath}
\usepackage{amssymb} 
\usepackage{siunitx}
\usepackage{graphicx}
\graphicspath{{./Figures/}}
\usepackage{stfloats}
\usepackage{breqn}
\usepackage{caption}
\usepackage{fancyhdr}
\usepackage{algorithm,algpseudocode}
\usepackage[noadjust]{cite}

\newcommand{\paramS}{$\theta$, $\omega$ and $\epsilon~$}
\newcommand{\paramEnd}{$\theta$, $\omega$ and $\epsilon$}
\newcommand{\paramVec}{$\boldsymbol{\theta},\boldsymbol{\omega},\boldsymbol{\epsilon}$}
\newcommand{\paramInEq}{\theta, \omega, \epsilon}
\newcommand{\paramInEqS}{\theta_{0:k}, \omega_{0:k}, \epsilon_{0:k}}
\newcommand{\paramInEqSI}{\theta_{0:k}^{(i)}, \omega_{0:k}^{(i)}, \epsilon_{0:k}^{(i)}}
\newcommand{\paramInEqSIK}{\theta_{0:k-1}^{(i)}, \omega_{0:k-1}^{(i)}, \epsilon_{0:k-1}^{(i)}}
\begin{document}
%
\title{Iterative Joint Parameters Estimation and Decoding in a Distributed Receiver for Satellite Applications and Relevant Cramer-Rao Bounds}

\author{Ahsan~Waqas,
		Khoa~Nguyen,
        	Gottfried~Lechner,
        and~Terence~Chan \\
{\textit{University of South Australia, Australia}}\\
{Email: \textit{ahsan.waqas@mymail.unisa.edu.au}}
}

%


\maketitle


\begin{abstract}
This paper presents an algorithm for iterative joint channel parameter (carrier phase, Doppler shift and Doppler rate) estimation and decoding of transmission over channels affected by Doppler shift and Doppler rate using a distributed receiver. This algorithm is derived by applying the sum-product algorithm (SPA) to a factor graph representing the joint {a posteriori} distribution of the information symbols and channel parameters given the channel output. In this paper, we present two methods for dealing with intractable messages of the sum-product algorithm. In the first approach, we use particle filtering with sequential importance sampling (SIS) for the estimation of the unknown parameters. We also propose a method for fine-tuning of particles for improved convergence. In the second approach, we approximate our model with a random walk phase model, followed by a phase tracking algorithm and polynomial regression algorithm to estimate the unknown parameters. We derive the Weighted Bayesian Cramer-Rao Bounds (WBCRBs) for joint carrier phase, Doppler shift and Doppler rate estimation, which take into account the prior distribution of the estimation parameters and are accurate lower bounds for all considered Signal to Noise Ratio (SNR) values. Numerical results (of bit error rate (BER) and the mean-square error (MSE) of parameter estimation) suggest that phase tracking with the random walk model slightly outperforms particle filtering. However, particle filtering has a lower computational cost than the random walk model based method.
\end{abstract}


%
\IEEEpeerreviewmaketitle

\section{Introduction}  
Software-defined radios (SDRs) have enabled a proliferation of small devices with wireless communication capabilities. For example, small sensors communicating their measurements via terrestrial or even satellite communication links. SDRs are versatile devices that allow sensors to send and receive signals in a wide range of frequency bands.

Long-range communication may be limited by small antenna gains. Cooperation between several sensors - also called nodes - enables communications to be established over long distances. By collaborating, nodes form a "virtual" high gain antenna that can form a beam in a desired direction for transmission (called distributed transmit beamforming (DTBF)) or reception (called distributed receive beamforming (DRBF)).

\begin{figure}
\centering
\centerline{\includegraphics[width=0.5\linewidth]{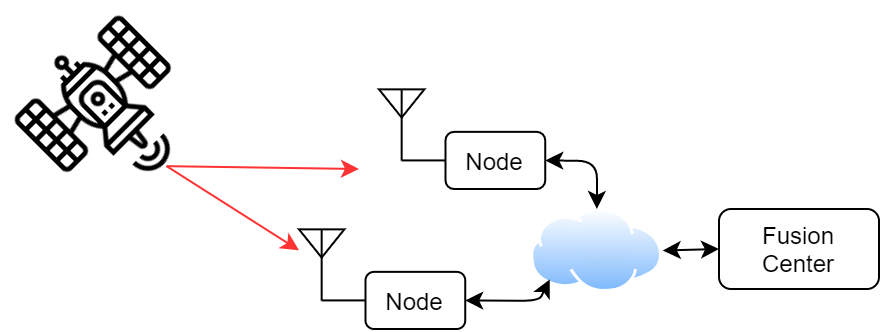}}
\caption{Distributed Receiver.}
\label{fig:DRBF}
\end{figure}

In this paper, we are interested in a distributed receiver as depicted in Fig.~\ref{fig:DRBF}, where $N$ distributed nodes can communicate with a fusion center. A scenario shown in Fig.~\ref{fig:DRBF} could correspond to a long-distance downlink in which multiple nodes collaborate to receive messages from a satellite or base station. Note that $N$ nodes can achieve maximum gain of $N$ by cooperation \cite{Quitin2}.

We assume that the backhaul links between the nodes and the fusion center facilitate bi-directional data exchange but do not support any reference signal for synchronization, such as wired and wireless network connections based on the IP protocol. This constraint is significantly different from research published recently, which focuses on dedicated links for synchronization \cite{Yan,Ouassal,Mghabghab}.


The majority of the literature assumes that signals are already synchronized at the receiving node, meaning that signals are sent to the fusion center without any time, phase or frequency offset \cite{Choi2,Bidigare,Brown}. In this paper, we focus on forwarding signals from each node to the fusion center when nodes are not synchronized in phase and frequency. Frequency offsets can occur for a variety of reasons, including imperfections in local oscillators and Doppler shifts due to the movement of the transmitter and receiver nodes. We can experience a change in frequency (Doppler rate) during a packet transmission as well, particularly in communications with low-earth orbit and medium-earth orbit satellites. The fusion center's main functions are to estimate and correct frequency and phase offsets in signals received from each node and to decode packets using all of the incoming signals. 

In general, phase and frequency can be estimated with blind or data-aided algorithms~\cite{Tianqi, Bofan, Tian, Huang}. The use of data-aided algorithms leads to a decrease in data transmission rates during packet transmissions, as they employ pilots. Though no initial training is required for blind schemes, feedback-based blind schemes like phase-locked loops (PLL) require a long convergence time; which cannot be achieved for small packets or bursty transmissions. Another type of algorithm is semi-data-aided (SDA), which uses both pilot and data symbols to estimate Doppler shift and Doppler rate in order to meet two requirements: accuracy of estimation and high spectral efficiency \cite{Yunpeng}. 

In the literature, a wide variety of algorithms are proposed for estimating the Doppler shift and Doppler rate that are non-iterative in nature \cite{Tianqi, Bofan, Tian, Huang}. In addition to non-iterative algorithms, numerous authors have developed blind and pilot-aided iterative algorithms based on factor graphs and the sum-product algorithm for single-channel transmission under phase noise \cite{Pecorino, Colavolpe, Shayovitz, Dauwels_phase}. In \cite{Alfredsson}, the authors consider iterative detection and correlated phase noise compensation for multi-channel optical transmission. A message passing algorithm for joint channel and phase-noise estimation for MIMO systems has been presented in \cite{Krishnan}. To the best of the author's knowledge, no prior work has been presented for semi data-aided algorithm for iterative joint phase, Doppler shift and Doppler rate estimation and decoding for the distributed multi-node receiver.  

In this paper, we employ message passing based algorithm for the iterative joint phase, Doppler shift and Doppler rate estimation and decoding. Based on both code constraints and channel statistics, this algorithm operates on a factor graph (FG) and uses the sum-product algorithm (SPA) to compute the messages.

Sum-product algorithms can produce messages containing intractable integrals. We present two methods to deal with intractable integrals. In the first approach, we consider the use of sequential Monte Carlo methods such as particle filtering, which has been applied to synchronization problems in communication systems \cite{Ahsan,NasirTime,Abdzadeh,Pedrosa}. The particle filter works by developing a recursive Bayesian filter to estimate the posterior joint probability of unknown parameters. In the second approach, we approximate our model with a random walk phase model, followed by a phase tracking algorithm and polynomial regression algorithm, which are used to estimate the unknown parameters (phase, Doppler shift, and Doppler rate).

To benchmark, we compare the estimation performance of the proposed algorithm with bounds. In the literature, different types of bounds are used under a different set of assumptions. In general, Cramer-Rao bounds (CRB) serve as benchmarks for unbiased estimators of parameters. The joint CRB for phase, Doppler shift and Doppler rate is derived in \cite{Giugno}. The joint CRBs, however, do not take into account the prior information about unknown parameters. A Bayesian CRB (BCRB) is a type of bound that incorporates the prior distribution of the unknown parameters. The Bayesian Cramer-Rao bound does not exist when parameters have uniform prior distributions \cite{van2007bayesian}. A weighted Bayesian Cramer-Rao (WBCRB) bound is appropriate in the case of uniform prior distributions. In this paper, we develop new WBCRBs for phase, Doppler shift and Doppler rate.

The major contributions of this paper are as follows:

\begin{enumerate}
\item We propose a sum-product message passing algorithm for joint phase, Doppler shift, Doppler rate estimation and decoding. 
\item We present two message computation methods based on particle filtering (with fine tuning the particle) and random-walk phase model. 
\item We derive the weighted Bayesian Cramer-Rao bounds of joint estimation of the phase, Doppler shift and Doppler rate.

\item Numerically, we demonstrate that weighted Bayesian Cramer-Rao bounds are valid bounds for parameter estimation when the parameters have uniform prior distributions. Furthermore, we demonstrate numerically that the fine-tuning of the particle filter significantly reduces the estimation error in comparison to a particle filter without fine-tuning. As a result of fine-tuning the particle filter with only 400 particles, the MSE approaches the CRB at SNR >3~dB. With only 100 phase quantization levels, the MSE of the random walk phase model approaches CRB at SNR> 2 dB. The proposed algorithm based on random walk model outperforms particle filtering technique in simulations of MSE and BER, however, it is more computationally intensive.
\end{enumerate}

\textit{Notations:} We use boldface capital letters to denote matrices. We denote sequences and vectors using boldface lowercase letters. In the short form, $\boldsymbol{x} = x_{0:k}$ denotes a sequence $\{x_0, x_1,\cdots,x_{k}\}$, while a matrix or vector is indicated by $[\cdot].$ A diagonal matrix is denoted by $\operatorname{diag}(\cdot).$ We denote transpose and Hermition of matrix by $[\cdot]^T$ and $[\cdot]^H$, respectively. $\boldsymbol{I}_L$ is a $L\times L$ identity matrix. $\mathbb{E}_{\theta}[\cdot]$ represents that expectation is performed with respect to the random variable $\theta$ and $\operatorname{Var}\left(\theta\right)$ represents the variance of $\theta.$ Furthermore, $z \sim \mathcal{U}\left(-z_m, z_m\right)$ represents that $z$ has a uniform distribution between $-z_m$ and $z_m$ and $z \sim \mathcal{CN}\left(0, \sigma^{2}\right)$ indicates that $z$ follows a complex Gaussian distribution with zero mean and variance $\sigma^{2}.$ We use notation $ z \sim \beta\left(U, V\right)$ to represent that $z$ has a beta distribution with shape parameters $U$ and $V.$ $ z \sim \mathcal{VM}(\mu, \kappa)$ depicts that $z$ has Tikhonov (von Mises) distribution.

\section{System Model} \label{sec:system_mode}  
In the following section, we first describe the signal model and then we discuss our estimation and decoding objective.
\subsection{Signal Model} 
\begin{figure*}
\centerline{\includegraphics[width=0.7\linewidth]{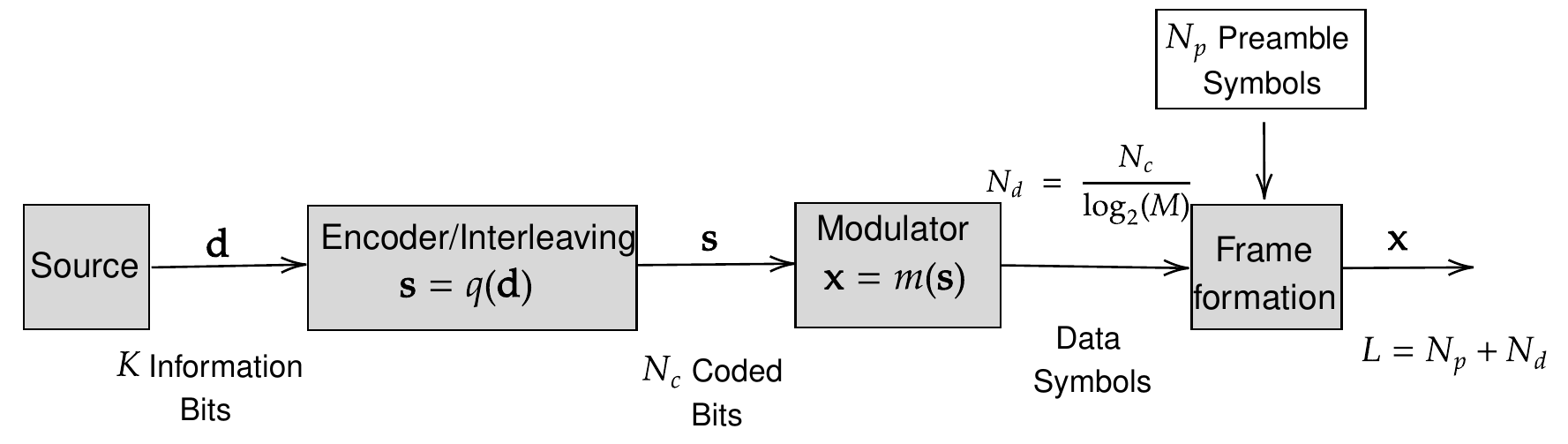}}
\caption{Block diagram of the transmitter.}
\label{fig:Tx_Block}
\end{figure*}

We consider a digital communication system where the transmitter transmits a sequence of $L$ complex modulated symbols $\boldsymbol{x} \overset{\Delta}{=} x_{0:L-1}$ over a noisy channel towards a distributed receiver. The channel input symbols sequence $\boldsymbol{x}$ is assumed to have $N_p$ preamble and $N_d$ information (data) symbols, so the total frame length is $L= N_p + N_d.$ We assume that both preamble and data symbols are mapped from the same complex modulation alphabet $\mathcal{X}$ of size $M.$ $N_c$ coded bits from a sequence $\boldsymbol{s}\overset{\Delta}{=} s_{1:N_c}$ are mapped to $N_d$ complex data symbols from $\mathcal{X}$, where $N_d = \frac{N_c}{\log_2(M)}$. While the coded bits $\boldsymbol{s}$ are derived following the encoding of information bits $\boldsymbol{d}\overset{\Delta}{=} d_{1:K}$ with a code rate of $R = K/N_c.$ The block diagram of the transmitter is shown in Fig.~\ref{fig:Tx_Block}, where $\boldsymbol{s}=q(\boldsymbol{d})$ and $\boldsymbol{x}=m(\boldsymbol{s})$ denote encoding and modulating functions, respectively.

The distributed receiver consists of $N_r$ receiving nodes. We assume an independent realization of time varying frequency offset (Doppler distortion) affects the received signal at each node. Furthermore, we assume perfect time synchronization is achieved at each node.  In addition, we assume a line-of-sight path between the transmitter and receive node without any relevant multipath propagation and hence we do not consider fading. The discrete time signal at $n^{th}$ receive node, after matched filtering and sampling at symbol rate, can be written as:
\begin{equation}\label{eq:channelmodel3}
y_{k,n}=x_{k} e^{j \phi_{k,n}}+v_{k,n} \quad \forall ~n \in \{1, 2, \cdots, N_r\}.
\end{equation}
Here $x_{k}$ is the channel input symbol at time $k \in \{0,1,2, \ldots, L-1\}$ and $y_{k,n}$ is the corresponding received sample at $n^{th}$ node.  $\phi_{k,n} \in (-\pi, \pi]$ is the unknown time-varying phase over the frame. $v_{k,n}$ is a complex white Gaussian noise realization with known variance $\sigma^{2},$ i.e., $\sigma^{2}/2$ per dimension. We assume that the additive noise is independent for each receive node. 

We assume that $\phi_{k,n}$ (given in \eqref{eq:channelmodel3}) is independent among receiving nodes. Based on the Taylor series expansion and by ignoring higher order terms, we can approximate $\phi_{k,n}$ with the polynomial of second degree as

\begin{equation}\label{eq:thetamodelQuadratic}
\begin{aligned}
\phi_{k,n}&= \left(\theta_{n} + \omega_{n} k+ \epsilon_{n} k^2 \right) \bmod 2 \pi \quad \forall ~n \in \{1, 2, \cdots, N_r\}.
\end{aligned}
\end{equation}
Here, $\theta_n$ is the initial phase, $\omega_{n} = 2\pi f_n T$ is the normalized frequency offset (Doppler shift) and $\epsilon_{n}= \pi f_{n}^{'} T^2$ is the normalized change in frequency (Doppler rate). For $n^{th}$ receive node, $f_n$, $f_{n}^{'}$ and $T$ denote frequency offset, change in frequency and symbol duration, respectively. The prior densities of \paramEnd \footnote{Throughout this paper, whenever we will drop index $n$ for estimation parameters \paramEnd, it is understood we are discussing parameters of one receive node as they are local to each receive node.} are assumed to be $\theta \sim \mathcal{U}(-\pi,\pi), ~\omega \sim \mathcal{U}(-\omega_m,\omega_m),$ and $\epsilon \sim \mathcal{U}(-\epsilon_m, \epsilon_m)$, respectively. Here, $\omega_m$ and $\epsilon_m$ denote the maximum possible value of Doppler shift and Doppler rate in the received signal, respectively. To avoid phase ambiguity, we assumed that $\omega_m \ll \frac{1}{T}$ and $\epsilon_m \ll \frac{1}{T^2}$.

\subsection{Estimation and Decoding Objective}
Let $\boldsymbol{y}_{n}=\left[y_{0, n}, \ldots, y_{L-1, n}\right]^{T} \in \mathbb{C}^{L \times 1}$ denote received samples of one frame at the $n^{th}$ receive node. Define the unknown parameters vectors for all receive nodes as $\boldsymbol{\theta}=\left[\theta_{1}, \ldots, \theta_{n}\right]^{T} \in (-\pi,\pi)^{N_r}$, $\boldsymbol{\omega}=\left[\omega_{1}, \ldots, \omega_{n}\right]^{T} \in (-\omega_m,\omega_m)^{N_r}$ and $\boldsymbol{\epsilon}=\left[\epsilon_{1}, \ldots, \epsilon_{n}\right]^{T} \in (-\epsilon_m, \epsilon_m)^{N_r}.$ Finally, let $\boldsymbol{y}\triangleq\left[\boldsymbol{y}_{1}, \boldsymbol{y}_{2}, \ldots, \boldsymbol{y}_{N_r}\right]^T$ denote all the received samples. In this paper, our objective is to develop an iterative algorithm which can jointly estimate the unknown parameters \paramVec~and the transmitted information bits $\boldsymbol{d}$ by taking into consideration all the received samples $\boldsymbol{y}$ of all the receive nodes. 

\section{Message Passing Using SPA}

The algorithm we shall discuss is called maximum {a posteriori} decoding (MAP), in which we estimate the transmitted information bits by taking advantage of the information available after observing all samples at all receiving nodes. According to the MAP algorithm, detection is performed bit by bit as
\begin{equation}\label{eq:x_MAP3L}\begin{aligned}
\hat{d}_{l}&=\operatorname*{arg\,max}_{d_{l} \in\{0,1\}}~ p\left(d_{l} \mid \boldsymbol{y}\right) \qquad \forall~l \in \{1,2,\cdots,K\},
\end{aligned}\end{equation}where $p\left(d_{l} \mid \boldsymbol{y}\right)$\footnote{Throughout this paper, by a slight abuse of notation, $p$ will denote probability density functions, probability mass functions and functions involving a mix of discrete and continuous variables. The arguments of $p$ will indicate which random variables are involved.} denotes the a posteriori probability mass function (PMF) of the $l^{th}$ information bit $d_{l}$ given the received samples $\boldsymbol{y}$. $p\left(d_{l} \mid \boldsymbol{y}\right)$ can be obtained by marginalizing the joint probability distribution function $p(d_{l}, \boldsymbol{\theta},\boldsymbol{\omega},\boldsymbol{\epsilon} |\boldsymbol{y})$ as 

\begin{equation}\label{eq:x_MAP4L}\begin{aligned}
p\left(d_{l} \mid \boldsymbol{y}\right)&= \int_{-\epsilon_m}^{\epsilon_m}\int_{-\omega_m}^{\omega_m}\int_{-\pi}^{\pi}  p(d_{l}, \boldsymbol{\theta},\boldsymbol{\omega},\boldsymbol{\epsilon} |\boldsymbol{y}) d\boldsymbol{\theta}~ d\boldsymbol{\omega}~ d\boldsymbol{\epsilon} 
\propto   \int_{-\epsilon_m}^{\epsilon_m}\int_{-\omega_m}^{\omega_m}\int_{-\pi}^{\pi} \sum_{\boldsymbol{d}\backslash \{d_l\}} p(\boldsymbol{d},\boldsymbol{y}, \boldsymbol{\theta},\boldsymbol{\omega},\boldsymbol{\epsilon}) d\boldsymbol{\theta}~ d\boldsymbol{\omega}~ d\boldsymbol{\epsilon} \\
\end{aligned}\end{equation}where $\propto$ denotes the proportionality and $\boldsymbol{d}\backslash \{d_l\}$ denotes all the elements of $\boldsymbol{d}$ except $d_l.$ We can further factorize $p(\boldsymbol{d},\boldsymbol{y}, \boldsymbol{\theta},\boldsymbol{\omega},\boldsymbol{\epsilon})$ as
\begin{equation}\label{eq:joint_pdf3L}
\begin{aligned}
p(\boldsymbol{d},\boldsymbol{y}, \boldsymbol{\theta},\boldsymbol{\omega},\boldsymbol{\epsilon})  &\overset{\text{(a)}}{=} p(\boldsymbol{y}| \boldsymbol{d}, \boldsymbol{\theta},\boldsymbol{\omega},\boldsymbol{\epsilon})p(\boldsymbol{d}| \boldsymbol{\theta},\boldsymbol{\omega},\boldsymbol{\epsilon})p( \boldsymbol{\theta},\boldsymbol{\omega},\boldsymbol{\epsilon})
\overset{\text{(b)}}{=}  p(\boldsymbol{y}| \boldsymbol{d}, \boldsymbol{\theta},\boldsymbol{\omega},\boldsymbol{\epsilon})p(\boldsymbol{d})p( \boldsymbol{\theta}) p(\boldsymbol{\omega}) p(\boldsymbol{\epsilon})\\
&\propto  p(\boldsymbol{d}) \prod_{n} p\left(\boldsymbol{y_{n}} \mid \boldsymbol{x}, \theta_n, \omega_n, \epsilon_n \right) p(\theta_n) p(\omega_n) p(\epsilon_n)\\
&\propto  p(\boldsymbol{d}) \prod_{n} \prod_{k} p\left(y_{k,n} \mid x_{k}, \theta_n, \omega_n, \epsilon_n \right) p(\theta_n) p(\omega_n) p(\epsilon_n),\\
\end{aligned}
\end{equation}where $\boldsymbol{x} \triangleq m(q(\boldsymbol{d}))$  indicates the transmitted sequence corresponding to $\boldsymbol{d}$. Further, (a) follows from the chain rule ${p(A,B) = p(B|A)p(A)}$ and (b) is obtained by using the assumption that information bits $\boldsymbol{d}$ and the unknown parameters (\paramVec) are independent from each other.

\subsection{Factor Graph (FG) Representation} \label{sec:FG_rep}

\begin{figure}
	\centering
	\includegraphics[width=0.4\linewidth]{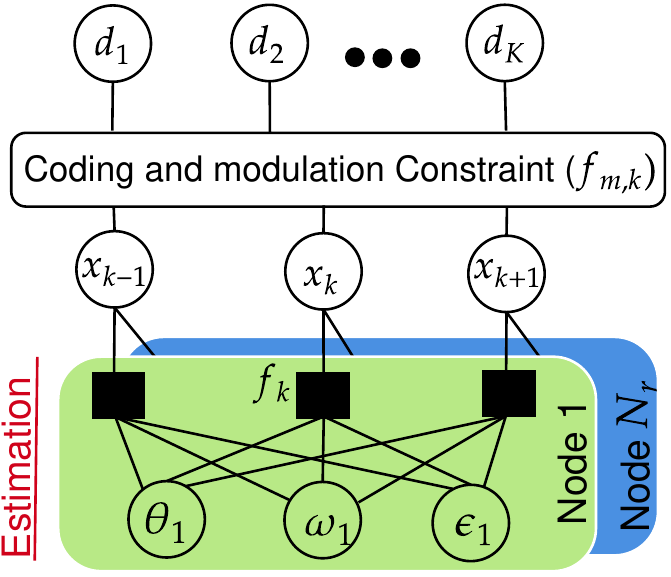}
  	\caption{Factor Graph of joint distribution function.}
  	\label{fig:joint_pdf3L}
\end{figure}

In digital communication, factor graphs are an important way of describing and constructing iterative message-passing algorithms. The function given in \eqref{eq:joint_pdf3L} can be represented by a factor graph shown in Fig.~\ref{fig:joint_pdf3L}. In this factor graph, a factor node is represented by a black square, whereas a variable node is represented by a circle. Each receive node has independent constant unknown parameters ($\theta_n$, $\omega_n$ and $\epsilon_n$) which is presented by a distinct color at the lower part of the factor graph\footnote{Note that "factor node" and "variable node" refer to components of a factor graph, whereas "receive node" refers to an independent receiver.}. As shown, FG can be broken into two parts. The upper portion of FG deals with demodulation, decoding and is common across all the receive nodes. Parameters estimation takes place in the lower portion of the graph, which is local to each receive node.

\subsection{Messages for Sum-Product Algorithm} \label{sec:SPA_messages}
Here, we discuss messages for factor graph based on sum-product algorithm, assuming the reader is familiar with sum-product algorithms. A complete discussion of messages for SPA is available in \cite{Kschischang}. In this paper, we use the notation $\mu_{start \rightarrow  end}(.)$ to represent message from a ${start}$ factor node/variable node towards an ${end}$ variable node/factor node. For example, $\mu_{f_k \rightarrow  x_k}(x_{k}) $ shows message from factor node $f_k$ towards variable node $x_k.$ \\

\subsubsection{Messages in the Lower Part of FG} The lower part of FG contains messages which are local to a receive node. Therefore, we drop subscript $n$ in this section. Inside the lower part of the FG, we have seven messages at the $k^{th}$ symbol. Three downward messages go from factor node $f_k$ towards unknown parameters variable nodes (\paramEnd) which are computed as 

\begin{equation}\label{eq:downMsg}
\begin{aligned}
\mu_{f_k \rightarrow  \theta}(\theta) &= \int_{-\epsilon_m}^{\epsilon_m}\int_{-\omega_m}^{\omega_m}\sum_{x_{k}} \mu_{x_k \rightarrow  f_k}(x_{k}) p(y_k \mid x_k, \theta,\omega,\epsilon) \mu_{\omega \rightarrow f_{k}}\left(\omega\right) \mu_{\epsilon \rightarrow f_{k}}\left(\epsilon\right)~d\omega~d\epsilon, \\
\mu_{f_k \rightarrow  \omega}(\omega) & = \int_{-\epsilon_m}^{\epsilon_m}\int_{-\pi}^{\pi}\sum_{x_{k}} \mu_{x_k \rightarrow  f_k}(x_{k}) p(y_k \mid x_k,\theta,\omega,\epsilon) \mu_{\theta \rightarrow f_{k}}\left(\theta\right) \mu_{\epsilon \rightarrow f_{k}}\left(\epsilon\right)~d\theta~d\epsilon, \\
\mu_{f_k \rightarrow  \epsilon}(\epsilon)&=\int_{-\omega_m}^{\omega_m}\int_{-\pi}^{\pi}\sum_{x_{k}} \mu_{x_k \rightarrow  f_k}(x_{k}) p(y_k \mid x_k,\theta,\omega,\epsilon) \mu_{\theta \rightarrow f_{k}}\left(\theta\right) \mu_{\omega \rightarrow f_{k}}\left(\omega\right) ~d\theta~d\omega,
\end{aligned}
\end{equation}where
$$p(y_k \mid x_k,\theta,\omega,\epsilon) \propto \exp \left\{-\frac{1}{ \sigma^{2}}\left|y_{k}- x_{k} e^{j \left(\theta + \omega k+ \epsilon k^2 \right)}\right|^{2}\right\}.$$The three upward messages from \paramS towards $f_k$ are computed using SPA as
\begin{equation}\label{eq:UpMsg}
\begin{aligned}
\mu_{\theta \rightarrow f_{k}}\left(\theta\right) &=\prod_{l, l \neq k} \mu_{f_l \rightarrow  \theta}(\theta)	\\
\mu_{\omega \rightarrow f_{k}}\left(\omega\right)&=\prod_{l, l \neq k} \mu_{f_l \rightarrow  \omega}(\omega)\\
\mu_{\epsilon \rightarrow f_{k}}\left(\epsilon\right) &=\prod_{l, l \neq k} \mu_{f_l \rightarrow  \epsilon}(\epsilon).
\end{aligned}
\end{equation}Finally, the upward message $\mu_{f_k \rightarrow  x_k}(x_{k})$ is computed as
\begin{equation}\label{eq:UpMsg_f_x}
\begin{aligned}
\mu_{f_k \rightarrow  x_k}(x_{k}) = \int_{-\epsilon_m}^{\epsilon_m}\int_{-\omega_m}^{\omega_m}\int_{-\pi}^{\pi}  &\mu_{\theta \rightarrow  f_k}(\theta) \mu_{\omega \rightarrow  f_k}(\omega)  \mu_{\epsilon \rightarrow  f_k}(\epsilon)p(y_k \mid x_k,\theta,\omega,\epsilon) ~d\theta ~ d\omega ~d\epsilon.
\end{aligned}
\end{equation}\\

\subsubsection{Messages in the Upper Part of FG} The upper part of FG is common for all nodes. In this section, the upward message from the lower part of FG, given in \eqref{eq:UpMsg_f_x}, will be denoted as $\mu_{f_{k,n} \rightarrow  x_k}(x_{k})$ for $n^{th}$ node. The message $ \mu_{x_{k} \rightarrow  f_{m,k}}(x_k)$ is computed by multiplying all the incoming messages from all receive nodes as
\begin{equation}\label{eq:UpMsg_x_fm}
\begin{aligned}
\mu_{x_{k} \rightarrow  f_{m,k}}(x_k) &=\prod_{n=1}^{N_r}\mu_{f_{k,n} \rightarrow  x_k}(x_{k}). 
\end{aligned}
\end{equation}

All other messages inside the upper part of factor graph are discussed in \cite{Pecorino, Colavolpe, Shayovitz, Dauwels_phase,Zhao2014}, which we do not discuss here. In summary, the upper part of FG takes PMF of the symbols as upward message from the lower part of FG and sends updated PMF of the symbols towards the lower part of FG. It is important to note that SPA messages for the lower part of the FG have intractable integrals, so to deal with this we will need to apply some approximation technique, which we will discuss in Section~\ref{sec:PF method} and \ref{sec:RW_model}.
\section{Particle Filter Method for Parameters Estimation}\label{sec:PF method}
In this section, we discuss particle filter based method to estimate the unknown parameters \paramS for one receive node.

\subsection{Particle Filter Formulation}

The lower part of the FG messages are local to each receive node which takes $\mu_{x_k \rightarrow  f_k}(x_{k})$ as input from the upper graph and sends out $\mu_{f_k \rightarrow  x_k}(x_{k})$ towards the upper part of the FG. The lower part of the factor graph shows the joint posterior probability density function (PDF) of the unknown parameters given the observations which is denoted as $p\left(\paramInEq \mid y_{0: L-1}, \mu_{x_k \to f_k}(x_k)\right).$ Here $y_{0: L-1}$ denotes receive samples and \paramS are unknown parameters for one receive node. This distribution is analytically intractable. As a result, we employ a particle filter, which recursively approximates $p\left(\paramInEq \mid y_{0: L-1}\right)$ with a discrete probability measure and random support. For time $k$, we define the discrete probability measure as \cite{Ghirmai:2005}

$$\Xi_{k} = \left\{\left(\paramInEqS \right)^{(i)},\mathrm{w}_{k}^{(i)}\right\}_{i=1}^{N}.$$
Here, $N$ shows the total number of particles. $\mathrm{w}_{k}^{(i)}$ is the weight of the $i^{th}$ particle at the $k^{th}$ time stamp and $\paramInEqSI$ are trajectories from time $0$ up to $k$ of $i^{th}$ particle. Note that \paramS represent constant unknown parameters and $\paramInEqSI$  are particles which do not represent time-varying parameters. The posterior PDF can be then approximated as

\begin{equation}\label{eq:psterior_pdf}
p\left(\paramInEq \mid y_{0: k}\right) \approx 
\sum_{i=1}^{N} \mathrm{w}_{k}^{(i)} \delta\left(\left( \theta_{k}, \omega_{k}, \epsilon_{k}\right)^{(i)}-\left(\paramInEq \right)\right),
\end{equation}where $\delta(\cdot)$ denotes Dirac's delta function. Approximation of posterior PDF in \eqref{eq:psterior_pdf} approaches $p\left(\paramInEq \mid y_{0: k}\right)$ as $N \to \infty.$

One of the most widely used techniques to implement the particle filter is the Sequential Importance Sampling (SIS) method. In this technique, $\Xi_{k}$ is recursively computed from $\Xi_{k-1}$ when the $k^{th}$ observation is available. The SIS technique uses an importance function to draw the new particles and appropriate corresponding weights are assigned to these particles for the recursive empirical approximation of the desired PDF \cite{tutorialPF}. According to SIS, the weights are defined as
\begin{equation}\label{eq:weight_1}
\mathrm{w}_{k}^{(i)} \propto \frac{p\left( \paramInEqSI \mid y_{0: k}\right)}{\pi\left( \paramInEqSI \mid y_{0: k}\right)}.
\end{equation}Note that, weight of a particle at $k^{th}$ time stamp depends on all the previous measures up to $k$. In \eqref{eq:weight_1}, $\pi(\cdot)$ is the importance function which is chosen to have the factorization of the form

\begin{equation}\label{eq:importance_pdf}
\begin{aligned}
\pi\left( \paramInEqSI \mid y_{0: k}\right)=  f_{A}\times f_{B}\times f_{C} \times \pi\left(\paramInEqSIK \mid y_{0: k-1}\right),\\
\end{aligned}
\end{equation}where
\begin{equation}\label{eq:importance_pdf_abc}
\begin{aligned}
f_{A} &= p\left(\theta_{k}^{(i)} \mid \paramInEqSIK, y_{0: k-1}\right) \\
f_{B} &= p\left(\omega_{k}^{(i)} \mid \paramInEqSIK, y_{0: k-1}\right) \\
f_{C} &= p\left(\epsilon_{k}^{(i)} \mid \paramInEqSIK, y_{0: k-1}\right). 
\end{aligned}
\end{equation}
$p(\cdot)$ can be factorized as follows (detailed steps are given in Appendix~\ref{Appendix1}).

\begin{equation}\label{eq:posterior_pdf_factor}
\begin{aligned}
&p\left(\paramInEqSI \mid y_{0: k}\right) \propto  f_{A}\times f_{B}\times f_{C} \times p\left(\paramInEqSIK \mid y_{0: k-1}\right) \times p\left(y_{k} \mid \theta_{k}^{(i)}, \omega_{k}^{(i)}, \epsilon_{k}^{(i)}\right).
\end{aligned}
\end{equation}Substituting \eqref{eq:posterior_pdf_factor} and \eqref{eq:importance_pdf} in \eqref{eq:weight_1} and after some manipulation, we obtain the recursive form of the weight as

\begin{equation}\label{eq:weight_2}
\mathrm{w}_{k}^{(i)}=\mathrm{w}_{k-1}^{(i)} p\left(y_{k} \mid  \theta_{k}^{(i)}, \omega_{k}^{(i)}, \epsilon_{k}^{(i)}\right).
\end{equation}We can write \eqref{eq:weight_2} with incoming message to lower part of FG after marginalization of $x_k$ as
\begin{equation}\label{eq:weight_3}
\begin{aligned}
\mathrm{w}_{k}^{(i)} &= \mathrm{w}_{k-1}^{(i)} \sum_{x_k} p(x_k) p\left(y_{k} \mid  x_k,\theta_{k}^{(i)}, \omega_{k}^{(i)}, \epsilon_{k}^{(i)}\right) =\mathrm{w}_{k-1}^{(i)} \sum_{x_k} \mu_{x_k \rightarrow  f_k}(x_{k})  p\left(y_{k} \mid  x_k,\theta_{k}^{(i)}, \omega_{k}^{(i)}, \epsilon_{k}^{(i)}\right).
\end{aligned}
\end{equation}

\subsection{Steps for SIS Particle Filtering} \label{sec:Algorithm}
This section discusses detailed implementation steps of SIS particle filter for joint parameters \paramS estimation and upward message computation. The algorithm consists of the following steps:\\
\subsubsection{Initialization} 
We assume that the prior distributions of the transmitted symbols and \paramS are known. The prior densities of the \paramS are ${\theta_{-1}\sim \mathcal{U}(-\pi,\pi)}$, $\omega_{-1}\sim \mathcal{U}(-\omega_m,\omega_m),$ and $\epsilon_{-1}\sim \mathcal{U}(-\epsilon_m,\epsilon_m)$, respectively. In addition, according to system model discussed in Section~\ref{sec:system_mode} due to preamble in the transmitted burst, we know the first $N_p$ symbols at each receive node. We initialize the weights of all the particles to be equal, i.e., $\mathrm{w}_{-1}^{(i)}=1 / N.$\\

\subsubsection{Importance Sampling} 
For the importance sampling from importance function in \eqref{eq:importance_pdf}, we obtain the \paramS samples from $f_{A},~f_{B}$ and $f_{C}$, respectively.
As none of $f_{A}$, $f_{B}$ and $f_{C}$ densities can be exactly determined, we use an importance function to get samples for $\theta_k$, $\omega_k$ and $\epsilon_k$. Two main requirements for choosing the importance function are 
\begin{enumerate}
\item The function domain must coincide with the desired PDF domain.
\item The function must be strictly positive \cite{tutorialPF}. 
\end{enumerate}

As $\theta_k$ is a circular quantity, we approximate $f_{A}$ by Tikhonov (von Mises) distribution, i.e., $\theta_k \sim \mathcal{VM}(\mu_{\theta,k}, \kappa_{\theta,k})$~\cite{Shayovitz}, where $\mu_{\theta,k}$ is the mean of $\theta_k$ which is given as 
$$
\mu_{\theta,k} = \angle{\sum_{i=1}^{N} \mathrm{w}_{k-1}^{(i)} e^{j {\theta}_{k-1}^{(i)} }}
$$and $\kappa_{\theta,k}$ is the shape parameter which is defined in~\cite{fisher_1993} (Section 4.5.5).
%

By following the approach in \cite{NasirTime}, we approximate $f_{B}$ and $f_{C}$ by a beta distribution. Since beta distribution has a range $[0,1],$ we translate particles of $\omega$ and $\epsilon$ as follows
\begin{equation}
\begin{aligned}
{\Omega}_{k}^{(i)} &= \frac{\omega_{k}^{(i)} + \omega_{m} }{2 \omega_{m}}, \qquad
{\varepsilon}_{k}^{(i)} &= \frac{\epsilon_{k}^{(i)} + \epsilon_{m} }{2 \epsilon_{m}}.\\
\end{aligned}
\end{equation}
We draw the samples of $\Omega$ and ${\varepsilon}$ from

\begin{equation}\label{eq:a_k}
\begin{aligned}
\Omega_{k} &\sim \beta\left(U_{\omega,k}, V_{\omega,k}\right) \qquad
\varepsilon_{k} &\sim \beta\left(U_{\epsilon,k}, V_{\epsilon,k}\right),
\end{aligned}
\end{equation}where beta distribution parameters $U_{k}$ and $V_{k}$ are obtained by~\cite{NasirTime}

\begin{equation}\label{eq:u_a_k}
\begin{array}{c}
U_{\omega,k}=\overline{\Omega}_{k}\left(\frac{\overline{\Omega}_{k}\left(1-\overline{\Omega}_{k}\right)}{\sigma_{\Omega_{k}}^{2}}-1\right) \\
V_{\omega,k}=\left(1-\overline{\Omega}_{k}\right)\left(\frac{\overline{\Omega}_{k}\left(1-\overline{\Omega}_{k}\right)}{\sigma_{\Omega_{k}}^{2}}-1\right),
\end{array}
\end{equation}where $\overline{\Omega}_{k}$ and $\sigma_{\Omega_{k}}^{2}$ can be computed as

\begin{equation}\label{eq:bar_a_k}
\begin{array}{c}
\overline{\Omega}_{k}=\sum_{i=1}^{N} \mathrm{w}_{k-1}^{(i)} \Omega_{k-1}^{(i)} \\
\sigma_{\Omega_k}^{2}=\sum_{i=1}^{N} \mathrm{w}_{k-1}^{(i)}\left(\Omega_{k-1}^{(i)}-\overline{\Omega}_{k}\right)^{2}.
\end{array}
\end{equation}$U_{\epsilon,k}$ and $V_{\epsilon,k}$ can be computed in a similar manner. After drawing $\Omega$ and $\varepsilon$ particles, we can translate them back to $\omega$ and $\epsilon$ respectively, i.e. $\omega_{k}^{(i)}=2\omega_{m} \Omega_{k}^{(i)} - \omega_{m}$. \\

\subsubsection{Weight Update}
After obtaining the new particles, we update their corresponding importance weights. We can write the weight update expression \eqref{eq:weight_3} as

\begin{equation}
\begin{aligned}
\tilde{\mathrm{w}}_{k}^{(i)} & \propto \mathrm{w}_{k-1}^{(i)} \sum_{X \in \mathcal{X}} \mu_{x_k \rightarrow  f_k}(x_{k}) p\left(y_{k} \mid x_{k}=X, {\theta}_{k}^{(i)}, {\omega}_{k}^{(i)}, {\epsilon}_{k}^{(i)}\right) 
\end{aligned}
\end{equation}where, $\tilde{\mathrm{w}}_{k}^{(i)}$ is the non-normalized importance weight for the $i^{th}$ particle and
 ${p\left(y_{k} \mid x_{k}=X, {\theta}_{k}^{(i)}, {\omega}_{k}^{(i)}, {\epsilon}_{k}^{(i)}\right) \propto \exp \left\{\frac{-|{y_k - \mu_{k}^{(i)}\left(X\right)|^2}}{\sigma^2}\right\}},$ where ${\mu_{k}^{(i)}\left(X\right)=X e^{j( ( \theta_{k}^{(i)} + \omega_{k}^{(i)} k + \epsilon _{k}^{(i)} k^2)\bmod 2 \pi)}}$.
Finally, we normalize the weights as
\begin{equation}\label{eq:normalize_weight}
\mathrm{w}_{k}^{(i)}=\frac{\tilde{\mathrm{w}}_{k}^{(i)}}{\sum_{n=1}^{N} \tilde{\mathrm{w}}_{k}^{(n)}}.
\end{equation}\\

\subsubsection{Resampling}
Degeneracy of particles is a well known problem in SIS algorithm implementation which means after a few time steps, most of the importance weights have negligible values $\left(\mathrm{w}_{k}^{(i)} \simeq 0\right)$. To address this problem, resampling of particles is performed whenever the effective sample size $N_{\mathrm{eff}}=\frac{1}{\sum_{i=1}^{N}\left(\mathrm{w}_{k}^{(i)}\right)^{2}}$ of the particle filter goes below a certain fraction of $N$ \cite{NasirTime}. In our algorithm, resampling is performed when $N_{\mathrm{eff}} \leq N/2.$
$N$ new particles are generated by sampling the discrete set $\left\{\left({\theta}_{k}, {\omega}_{k}, {\epsilon}_{k}\right)^{(i)}\right\}_{i=1}^{N}$ with probabilities $\mathrm{w}_{k}^{(i)}$ and then resetting the importance weights to equal values $1 / N.$ \\

\subsubsection{Mean Calculation}
Finally, the importance weights and the drawn particles for the unknown parameters are used to compute the means of $\hat{\omega}_{k}$ and $\hat{\epsilon}_{k}$ as

\begin{equation}
\begin{aligned}
\hat{\omega}_{k}&= \sum_{i=1}^{N} {\omega}_{k}^{(i)} \mathrm{w}_{k}^{(i)},\quad
\hat{\epsilon}_{k}&= \sum_{i=1}^{N} {\epsilon}_{k}^{(i)} \mathrm{w}_{k}^{(i)}. \\
\end{aligned}
\end{equation}As $\theta$ is a circular quantity, we compute weighted circular mean to estimate $\hat{\theta}_{k} = \angle{r}$, where
\begin{equation}
\begin{aligned}
r &= \sum_{i=1}^{N} \mathrm{w}_{k}^{(i)} e^{j {\theta}_{k}^{(i)} }, \quad
\end{aligned}
\end{equation}and unwrapped mean phase at time $k$ is computed as
 \begin{equation}\label{eq:phi_m}
\hat{\phi}_{k} = \hat{\theta}_{k} + \hat{\omega}_{k} k + \hat{\epsilon}_{k} k^2.
\end{equation}\\

\subsubsection{Upward Message $(\mu_{f_k \rightarrow  x_k}(x_{k}) )$}
The upward message $\mu_{f_k \rightarrow  x_k}(x_{k}) $ is a PMF of the symbol at $k^{th}$ time index. We can compute upwards message in online mode (forward mode) using $\hat{\phi}_{k}$ from \eqref{eq:phi_m} as
\begin{equation}\label{eq:msg_fk_xk}
\mu_{f_k \rightarrow  x_k}(x_{k}=X_m) = \frac{p(y_k \mid \hat{\phi}_{k},x_k=X_m) }{\sum_{X \in \mathcal{X}} p(y_k \mid \hat{\phi}_{k},x_k=X) },
\end{equation}where $p(y_k \mid\hat{\phi}_{k},x_k) \propto \exp \left\{\frac{-1}{\sigma^2}|y_k - x_k e^{j \hat{\phi}_{k}}|^2\right\}.$ The computed message $\mu_{f_k \rightarrow  x_k}(x_{k}=X_m)$ in \eqref{eq:msg_fk_xk} does not incorporate future symbols for message computation. We propose that at the end of frame ($k=L-1$), when we have made use of all the available information to estimate $\theta$, $\omega$ and $\epsilon$, the upward messages can be computed in parallel for $k \in \{N_p,N_p+1,\cdots,L-1\}$ as
\begin{equation}\label{eq:msg_fk_xk2}
\mu_{f_k \rightarrow  x_k}(x_{k}=X_m) = \frac{p(y_k \mid \tilde{\phi}_{k},x_k=X_m) }{\sum_{X \in \mathcal{X}} p(y_k \mid \tilde{\phi}_{k},x_k=X) },
\end{equation}where
$$ 
\tilde{\phi}_{k} = \hat{\theta}_{L-1} + \hat{\omega}_{L-1} k + \hat{\epsilon}_{L-1} k^2 \qquad \forall k \in \{N_p,N_p+1,\cdots,L-1\}.
$$The proposed algorithm is summarized as Algorithm~\ref{alg:two}.\\
\begin{algorithm}[t]
\caption{Proposed Algorithm with SIS Particle Filter}\label{alg:two}
\begin{algorithmic}
\For{$iter = 1$ to $G$ (Total number of iterations between estimation and decoding)}
	\For{$n = 1$ to $N_r$ (Total number of nodes)}
		\State Initialize 
		$\theta_{-1}\sim \mathcal{U}(-\pi,\pi),\omega_{-1}\sim \mathcal{U}(-\omega_m,\omega_m),$ and $				\epsilon_{-1}\sim \mathcal{U}(-\epsilon_m,\epsilon_m)$
		\State $\mathrm{w}_{-1}^{(i)}=1 / N \quad \forall~i=1,2, \ldots, N$

		\For{$k = 0$ to $L-1$ (Total number of symbols)}
    			\State Compute $\overline{\Omega}_{k}, \overline{\varepsilon}_{k}, \sigma_{\Omega_k}^{2}$ and 				$\sigma_{\varepsilon_k}^{2}$ using \eqref{eq:bar_a_k} 
    			\State Compute $\mu_{\theta,k}, U_{\omega,k}, U_{\epsilon,k}, \kappa_{\theta,k}, V_{\omega,k}$ 				and $V_{\epsilon,k}$ using \eqref{eq:u_a_k}
    			\State Draw $N$ samples of $\theta_{k} \sim \mathcal{VM}\left(\mu_{\theta,k}, \kappa_{\theta,k}\right)$,  $					\Omega_{k} \sim \beta\left( U_{\omega,k}, V_{\omega,k}\right)$ and $\varepsilon_{k} \sim 						\beta\left(U_{\epsilon,k}, V_{\epsilon,k}\right)$
    			
			\For{$i = 1$ to $N$ (Total number of particles)}
         			\State back translation of particles $\omega_{k}^{(i)}=2\omega_{m} \Omega_{k}^{(i)} - 							\omega_{m}$ (similarly for $\epsilon$)
         			\State $\tilde{\mathrm{w}}_{k}^{(i)} = \mathrm{w}_{k-1}^{(i)} \sum_{X \in \mathcal{X}} \mu_{x_k \rightarrow  f_k}(x_{k}=X) p\left(y_{k} \mid x_{k}=X, {\theta}_{k}^{(i)}, {\omega}_{k}^{(i)}, {\epsilon}_{k}^{(i)}\right)  $
    			\EndFor

    			\State Normalize weights $\mathrm{w}_{k}^{(i)}=\frac{\tilde{\mathrm{w}}_{k}^{(i)}}{\sum_{l=1}^{N} \tilde{\mathrm{w}}_{k}^{(l)}}$
    			\State Resample if $ N_{\mathrm{eff}}=\frac{1}{\sum_{i=1}^{N}\left(\mathrm{w}_{k}^{(i)}\right)^{2}} \leq N / 2$
   			 \If { $ k > N_p$ and $\sigma_{\vartheta_k}^{2}< \theta_{\text{th}}$ and $\sigma_{\Omega_k}^{2}< \omega_{\text{th}}$}
           		\State Fine-tuning of particles as discussed in Section \ref{sec:FineTune}
         		\EndIf
    \State Estimation of parameters \quad
    $\hat{\omega}_{k} = \sum_{i=1}^{N} {\omega}_{k}^{(i)} \mathrm{w}_{k}^{(i)}, \quad
    \hat{\epsilon}_{k} = \sum_{i=1}^{N} {\epsilon}_{k}^{(i)} \mathrm{w}_{k}^{(i)}, \quad \hat{\theta}_{k} = \angle{\{\sum_{i=1}^{N} \mathrm{w}_{k}^{(i)} e^{j {\theta}_{k}^{(i)} }\}}$
    \State $\hat{\phi}_{k} = \hat{\theta}_{k} + \hat{\omega}_{k} k+ \hat{\epsilon}_{k} k^2$
\EndFor
\State Compute $\mu_{f_{k,n} \rightarrow  x_k}(x_{k})$ using \eqref{eq:msg_fk_xk2} $\forall~ k \in \{N_p,N_p+1,\cdots,L-1\}$
\EndFor
\State Compute internal message of modulation and coding constraints and compute messages upto $\mu_{x_k \rightarrow  f_k}(x_{k}) $
\EndFor
\end{algorithmic}
\end{algorithm}

\subsection{Fine-tuning (FT) of Particles}\label{sec:FineTune}

The lack of dynamics of constant parameters leads the particle filter to quickly converge to potentially wrong estimate after a few time instances, and when the particle filter algorithm has not explored the sample space~\cite{Kantas}. In our case, because the quadratic phase is only dominating at large value of $k$, the particles of $\epsilon$ usually quickly converge to a wrong value. We propose the following fine-tuning algorithm to prevent the particles from converging to a wrong value and to improve the estimation of $\epsilon.$

\begin{algorithm}
\caption{Proposed method for fine-tuning of particles}\label{alg:Fine_tuning}
\begin{algorithmic}
\State $m = 0$ to $k-1$
\State $f(m) = \tilde{\theta} + \tilde{\omega}m + \tilde{\epsilon}m^{2} ~\forall m$
\State Find $\tilde{\theta}, \tilde{\omega}$ and $\tilde{\epsilon}$ by using least-square method for quadratic curve fitting between $f(m)$ and $\hat{\phi}_m$
\For{$i = 1$ to $N$ (Total number of particles)}
         \State $\theta_k^{(i)} = \tilde{\theta} + \mathrm{rand}(-\alpha,\alpha)$
         \State $\omega_k^{(i)} = \tilde{\omega} + \mathrm{rand}(-\zeta,\zeta)$
         \State $\epsilon_k^{(i)} = \tilde{\epsilon} + \mathrm{rand}(-\gamma,\gamma)$
         \State $\tilde{\mathrm{w}}_{k}^{(i)} \propto \sum_{X \in \mathcal{X}} \mu_{x_k \rightarrow  f_k}(x_{k}) p\left(y_{k} \mid x_{k}=X_{m}, {\theta}_{k}^{(i)}, {\omega}_{k}^{(i)}, {\epsilon}_{k}^{(i)}\right)  $
   \EndFor
 \State    Normalize weights $\mathrm{w}_{k}^{(i)}=\frac{\tilde{\mathrm{w}}_{k}^{(i)}}{\sum_{l=1}^{N} \tilde{\mathrm{w}}_{k}^{(l)}}$\\
Note:~rand function generates random numbers in the range given in parenthesis by using uniform distribution.
\end{algorithmic}
\end{algorithm}

The proposed fine-tuning starts when the particles of $\theta$ and $\omega$ have converged at the $k^{th}$ symbol ($k > N_p$); such that variances of particles of $\theta$ and $\omega$ are under certain thresholds (call them $\theta_{\text{th}}$ and $\omega_{\text{th}}$). Then, we fit unwrapped estimated phase $\hat{\phi}_{m}$ with the curve $f(m) = \tilde{\theta} + \tilde{\omega}m + \tilde{\epsilon}m^{2} ~ \forall m \in \{0,1,\cdots,k-1\}$. Here, the unwrapped estimated phase $\hat{\phi}_{m}$ at each observation can be computed as given in \eqref{eq:phi_m}. We use quadratic curve fitting least square method to get the values of $\tilde{\theta}, \tilde{\omega}$ and $\tilde{\epsilon}$. We regenerate the particles of $ \theta, ~\omega$ and $\epsilon$ in the range $\tilde{\theta}\pm \alpha$, $\tilde{\omega}\pm \zeta$ and $\tilde{\epsilon} \pm \gamma$, respectively and update the corresponding weights. Here, $\alpha$, $\zeta$ and $\gamma$ are design parameters and they can be chosen to have a minimal residual error in parameters estimation. The proposed fine-tuning of particles is given in Algorithm~\ref{alg:Fine_tuning}.\\

\section{Random Walk (RW) Model Based Approximation}\label{sec:RW_model}

In this section, we discuss an indirect approach to estimate the unknown parameters \paramS for one receive node. As the phase model given in \eqref{eq:thetamodelQuadratic} is time-varying over one frame, we can use any phase tracking algorithm to estimate the phase at each symbol time. We can then estimate \paramS by applying a quadratic polynomial fit on all unwrapped estimated phases over one frame. For this purpose, we model the time-varying phase $\phi_{k} \in (-\pi, \pi]$ as a Wiener process given as
\begin{equation}\label{eq:thetamodelRW}
\phi_{k}=\left(\phi_{k-1}+W_{k}\right) \bmod 2 \pi,\end{equation}
where $W_{k}$ is white Gaussian noise with variance $\sigma_{W}^{2}.$ The phase model given in \eqref{eq:thetamodelRW} can track phase of model \eqref{eq:thetamodelQuadratic} when an appropriate value of $\sigma_{W}^{2}$ is used. We can find a reasonable value of $\sigma_{W}^{2}$ by using the prior distributions of $\omega$ and $\epsilon$ parameters. For phase model \eqref{eq:thetamodelQuadratic}, the phase change $\delta \phi$ between two consecutive symbols will be the largest at the end of the frame because phase model \eqref{eq:thetamodelQuadratic} has terms involving symbol index $k.$ The largest change in phase at the end of frame is ${\delta \phi_{m} = \omega_{m} + \epsilon_{m}((L-1)^2 - (L-2)^2) = \omega_{m} + \epsilon_{m}(2L-3)}.$ Now, by assuming that phase change between two consecutive symbols is uniformly distributed, i.e., $\delta \phi \sim \mathcal{U}(-\delta \phi_{m}, \delta \phi_{m}),$ we approximate $\sigma_{W}^{2} = \frac{1}{12}2\delta \phi_{m}= \frac{1}{6}\delta \phi_{m}$ for phase model \eqref{eq:thetamodelRW}. In short, phase model given in \eqref{eq:thetamodelRW} can track phase of model \eqref{eq:thetamodelQuadratic} when $\sigma_{W}^{2} = \frac{1}{6}\delta \phi_{m}.$ 

\begin{figure}
	\centering
  	\includegraphics[width=0.4\linewidth]{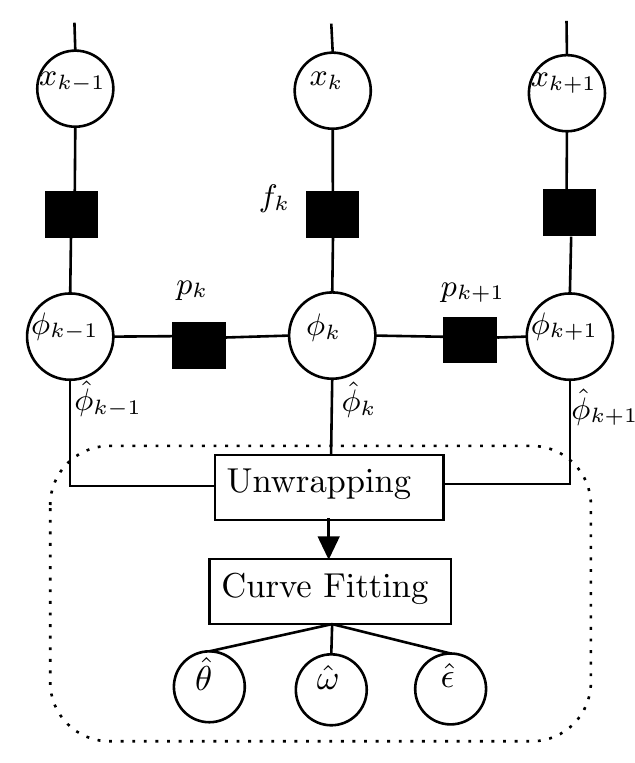}
  	\caption{Random walk phase model followed by unwrapping and curve fitting.}
\label{fig:RW_phase}
\end{figure}

The factor graph with random walk phase model followed by unwrapping of phase and curve fitting is shown in Fig.~\ref{fig:RW_phase}. We have summarized SPA messages in Table~\ref{tab:SPA_MSG_RW} and more detailed discussion about SPA messages for random walk phase model can be found in \cite{Shachar, Wang, Qiaolin, Kreimer}. The integral in $\mu_{p_{k} \rightarrow \phi_{k}}\left(\phi\right) $ is intractable. This integral can be computed (with reasonable computation cost) using rectangular integration rule as it involves integration only over one parameter $\phi$. By this method, integral is replaced by finite sum. Consider $\phi \in (-\pi, \pi]$ is quantized into $N_q$ levels, then $\mu_{p_{k} \rightarrow \phi_{k}}\left(\phi\right)$ can be written as

\begin{equation}\label{eq:f-d22}\begin{aligned}
\mu_{p_{k} \rightarrow \phi_{k}}\left(\frac{2\pi \ell}{N_q}\right)&=\sum_{m=0}^{N_q-1}  \mu_{\phi_{k-1} \rightarrow p_{k}}\left(\frac{2\pi m}{N_q}\right) p_{k}\left(\frac{2\pi \ell}{N_q} \mid \frac{2\pi m}{N_q}\right) \qquad \forall \ell \in \{0,1,\cdots,N_q-1\}
\end{aligned}\end{equation}where
\begin{equation}p_{k}\left(\phi_{k} \mid \phi_{k-1}\right) \triangleq\left(2 \pi \sigma_{W}^{2}\right)^{-1 / 2} \sum_{g \in \boldsymbol{Z}} e^{-\left(\left(\phi_{k}-\phi_{k-1}\right)+ g 2 \pi\right)^{2} / 2 \sigma_{W}^{2}}.
\end{equation}Here, $\boldsymbol{Z}$ is a set of integers. This approach is equivalent of applying SPA to a quantized phase model \cite{Dauwels_phase}. The message from variable node $\phi_k$ towards unwrapping function is the mean value of phase $\hat{\phi}_k$ which is computed as
\begin{equation}
\hat{\phi}_k = \sum_{m=0}^{N_q-1} \frac{2\pi m}{N_q} \mu_{p}(2\pi m/N_q) ,
\end{equation} where
\begin{equation}\label{eq:u_p}
\begin{aligned}
\tilde{\mu}_{p}(\phi) &= \mu_{f_k \rightarrow  \phi_k}(\phi)  \mu_{p_k \rightarrow  \phi_k}(\phi)  \mu_{p_{k+1} \rightarrow  \phi_k}(\phi) \\
\mu_{p}(\phi) &=\frac{\tilde{\mu}_{p}(\phi) }{\sum_\phi \tilde{\mu}_{p}(\phi) }.
\end{aligned}\end{equation}

Next, we fit unwrapped estimated phase $\hat{\phi}_k$ with the curve $f(k) = \hat{\theta} + \hat{\omega}k + \hat{\epsilon}k^{2} ~\forall k \in \{0,1, \cdots, L-1\}$. We use quadratic curve fitting least square method to get the values of $\hat{\theta}, \hat{\omega}$ and $\hat{\epsilon}$. After curve fitting, we compute the upward message as
\begin{equation}
\begin{aligned}
\mu_{f_k \rightarrow  x_k}(x_{k}) &= \exp \left\{\frac{-1}{\sigma^2}|y_k - x_k e^{j (\hat{\theta} + \hat{\omega} k + \hat{\epsilon} k^2)}|^2\right\} \qquad \forall k \in \{N_p,N_p+1,\cdots,L-1\}.
\end{aligned}
\end{equation}
Note that we do not compute $\mu_{\phi_k \rightarrow f_k}(\phi)$ and  $\mu_{f_k \rightarrow  x_k}(x_{k})$ directly from SPA, since this approach does not make use of the quadratic model. Instead, we first estimate $\hat{\theta}, \hat{\omega}$ and $\hat{\epsilon}$ and then use to compute the message  $\mu_{f_k \rightarrow  x_k}(x_{k}).$

\begin{center}
\begin{table}%
\centering
\caption{SPA Messages for Random Walk Phase Model}
\begin{tabular}{ |p{2cm}|p{6cm}|  }
 \hline
 Message 				 			&Computation						\\
 \hline
 $\mu_{f_k \rightarrow  \phi_k}(\phi)$			& $\sum_{x_{k}} \mu_{x_k \rightarrow  f_k}(x_{k})\exp \left\{-\frac{1}{ \sigma^{2}}\left|y_{k}- x_{k} e^{j \phi_k}\right|^{2}\right\}$			\\
\hline
 $\mu_{p_{k} \rightarrow \phi_{k}}\left(\phi\right) $			& $\int_{-\pi}^{\pi} \mu_{\phi_{k-1} \rightarrow p_{k}}\left(\phi \right) p_{k}\left(\phi_{k} \mid \phi_{k-1}\right) d\phi_{k-1}$\\
\hline
$\mu_{\phi_{k} \rightarrow p_{k+1} }\left(\phi\right) $  		&$\mu_{p_{k} \rightarrow \phi_{k}}\left(\phi\right) \mu_{f_{k} \rightarrow \phi_{k}}(\phi)$								\\
\hline
$\mu_{\phi_{k} \rightarrow p_{k} }\left(\phi\right) $  		 &$ \mu_{p_{k+1} \rightarrow \phi_{k}}\left(\phi\right) \mu_{f_{k} \rightarrow \phi_{k}}(\phi)$								\\
\hline
 \end{tabular}
\label{tab:SPA_MSG_RW}
 \end{table}
\end{center}

\begin{algorithm}
\caption{Proposed Algorithm with Random Walk Phase Model and Curve Fitting}\label{alg:Random_walk}
\begin{algorithmic}
\For{$iter = 1$ to $G$ (Total number of iterations between estimation and decoding)}
\For{$n = 1$ to $N_r$ (Total number of nodes)}
\State Initialize $\mu_{p_{0} \rightarrow  \phi_{0}}(\phi) = 1/N$ and $\mu_{p_{L} \rightarrow  \phi_{L-1}}(\phi) = 1/N$
\State $\phi = 1:\frac{2\pi}{N}:2\pi$\\
\For{$k = 0$ to $L-1$}\qquad\qquad \% Forward recursion 
\State Update $\mu_{p_{k} \rightarrow  \phi_{k}}(\phi)$, $\mu_{f_{k} \rightarrow  \phi_{k}}(\phi)$ and $\mu_{\phi_{k} \rightarrow p_{k+1} }\left(\phi\right)$
\EndFor\\
\For{$k = L-1$ to $0$}\qquad\qquad \% Backward recursion
\State Update $\mu_{p_{k+1} \rightarrow \phi_{k}}\left(\phi\right)$ and  $\mu_{\phi_{k} \rightarrow p_{k-1} }\left(\phi\right)$ 
\EndFor
\State Compute $\mu_{p}(\phi)$ using \eqref{eq:u_p}
\State Compute $\hat{\phi}_k = \sum_{m=0}^{N-1} \frac{2\pi m}{N} \mu_{p}(2\pi m/N) ~\forall k \in \{0,1,\cdots,L-1\}$
\State Unwrap the phases
\State Quadratic curve fit and estimate $\hat{\theta}, \hat{\omega}$ and $\hat{\epsilon}$
\State update upward message ${\mu_{f_k \rightarrow  x_k}(x_{k}) = \exp \left\{\frac{-1}{\sigma^2}|y_k - x_k e^{j (\hat{\theta}_{k} + \hat{\omega}_{k} k + \hat{\epsilon}_{k} k^2)}|^2\right\}}$
\EndFor
\State Compute $\mu_{x_{k} \rightarrow  f_{m,k}}(x_k)= \prod_{n=1}^{N_r}\mu_{f_{k,n} \rightarrow  x_k}(x_{k})$
\State Compute internal message of modulation and coding constraints and compute messages upto $\mu_{x_k \rightarrow  f_k}(x_{k}) $
\EndFor
\end{algorithmic}
\end{algorithm}
\section{Computational Complexity}\label{sec:Complexity}%

This section discusses the computational complexity of the proposed methods. For the particle filter method, we assume that the computational cost of drawing one multidimensional sample from a beta distribution or Tikhonov (von Mises) distribution is $\mathcal{O}(N_s^2)$ \cite{DURAN2012317}. Here $N_s$ is the dimension of the sample (number of unknown parameters) and in our case $N_s = 3.$ The complexity of particle filter for a single iteration (at one symbol) is dominated by importance sampling with the computational cost of $\mathcal{O}(N N_s^2),$ where $N$ is the number of particles \cite{DURAN2012317}. 

The complexity of the random-walk based method for a single iteration (at one symbol) is $\mathcal{O}(N_q^2)$ with naive implementation \cite{Colavolpe}, which is clear from the message given in \eqref{eq:f-d22}. Here $N_q$ is the number of phase quantization levels. Note that, we can improve implementation complexity to $\mathcal{O}(N_q \log N_q)$ by performing Toeplitz matrix multiplication using the Fast Fourier transform \cite{Beliakov2014OnFM}.

Since both proposed methods (random-walk and particle filter with fine-tuning) perform curve fitting, both methods are equally complex in this respect. For a random-walk based method, the only additional complexity is added during the phase unwrapping. Note that, for particle filter method we do not perform unwrapping on phase. 

In summary, the particle filter method scales linearly with number of particles and random-walk method scales non-linearly with phase quantization levels. Both method scales linearly with number of symbols and number of nodes. We must emphasize that this method of quantifying complexity only provides an approximate view of the hardware requirements for implementing the algorithms. As a result, it serves as a starting point for a more detailed analysis, which is beyond the scope of this paper.


\section{Data-Aided Cramer-Rao Bounds}\label{sec:CRBounds}

As \paramS are local parameters of each receive node, we discuss two types of data-aided Cramer-Rao bounds for a single receive node in this section. Firstly, we discuss the joint Cramer-Rao Bounds (JCRB) of \paramEnd. We then derive weighted Bayesian Cramer-Rao bounds (WBCRB) using prior information of \paramEnd.\\

\subsection{Joint Cramer-Rao Bound (JCRB)}
For a single receive node, we can write our signal model from \eqref{eq:channelmodel3} as
\begin{equation}\label{eq:channelmodelQuadratic}
\boldsymbol{y} = \boldsymbol{A B C x}+\boldsymbol{v},
\end{equation}where $\boldsymbol{y}=\left[y_{0}, y_{1}, \ldots, y_{L-1}\right]^{T}, \boldsymbol{v}=\left[v_{0}, v_{1}, \ldots, v_{L-1}\right]^{T}, \boldsymbol{x}=\left[{x}_{0}, {x}_{1}, \ldots, {x}_{L-1}\right]^{T}$ , $\boldsymbol{A}= e^{j \theta} \boldsymbol{I}_L$ and the diagonal matrices, ${\boldsymbol{B}=\operatorname{diag}\left(e^{j \omega(0)}, e^{j \omega(1)}, \ldots, e^{j \omega(L-1)}\right)}$, ${\boldsymbol{C}=\operatorname{diag}\left(e^{j \epsilon(0)}, e^{j \epsilon(1)}, e^{j \epsilon(4)}, \ldots, e^{j \epsilon(L-1)^2}\right)}.$ To derive data-aided bounds, we assume that $\boldsymbol{x}$ is known.

Let $\boldsymbol{\psi} \triangleq[\theta,\omega,\epsilon]^{T}$ and denote $\boldsymbol{\mu}= \boldsymbol{A} \boldsymbol{B} \boldsymbol{C}\boldsymbol{x}.$ Since the covariance matrix $\boldsymbol{K}_{y}$ does not depend on $\boldsymbol{\psi} ,$ the Fisher Information matrix (FIM) is 
\begin{equation} \label{eq:FIM}
\boldsymbol{J}(\boldsymbol{\psi}) \overset{\Delta}{=} 2 \Re\left(\frac{\partial \boldsymbol{\mu}^{H}}{\partial \boldsymbol{\psi}} \boldsymbol{K}_{y}^{-1} \frac{\partial \boldsymbol{\mu}}{\partial \boldsymbol{\psi}^{T}}\right),
\end{equation}
where $\frac{\partial}{\partial \boldsymbol{\psi}}(\cdot)$ denotes the partial derivative with respect to $\boldsymbol{\psi} ,$ the inverse covariance matrix, $\boldsymbol{K}_{y}^{-1}=\frac{1}{\sigma^{2}}\boldsymbol{I}_{L}$. Substituting in \eqref{eq:FIM}, we get

\begin{equation}\label{eq:FIM2}
\begin{aligned}
\boldsymbol{J}(\boldsymbol{\psi}) &=\frac{2}{\sigma^{2}} \Re\left(\frac{\partial \boldsymbol{\mu}^{H}}{\partial \boldsymbol{\psi}} \frac{\partial \boldsymbol{\mu}}{\partial \boldsymbol{\psi}^{T}}\right) =\frac{2}{\sigma^{2}} \Re\left(\left[\begin{array}{ccc}
\frac{\partial \boldsymbol{\mu}^{H}}{\partial \theta} \frac{\partial \boldsymbol{\mu}}{\partial \theta} & \frac{\partial \boldsymbol{\mu}^{H}}{\partial \theta} \frac{\partial \boldsymbol{\mu}}{\partial \omega}  & \frac{\partial \boldsymbol{\mu}^{H}}{\partial \theta} \frac{\partial \boldsymbol{\mu}}{\partial \epsilon}\\ \\%
\frac{\partial \boldsymbol{\mu}^{H}}{\partial \omega } \frac{\partial \boldsymbol{\mu}}{\partial \theta} & \frac{\partial \boldsymbol{\mu}^{H}}{\partial \omega} \frac{\partial \boldsymbol{\mu}}{\partial \omega} & \frac{\partial \boldsymbol{\mu}^{H}}{\partial \omega} \frac{\partial \boldsymbol{\mu}}{\partial \epsilon} \\\\%
\frac{\partial \boldsymbol{\mu}^{H}}{\partial \epsilon} \frac{\partial \boldsymbol{\mu}}{\partial \theta} & \frac{\partial \boldsymbol{\mu}^{H}}{\partial \epsilon} \frac{\partial \boldsymbol{\mu}}{\partial \omega} & \frac{\partial \boldsymbol{\mu}^{H}}{\partial \epsilon} \frac{\partial \boldsymbol{\mu}}{\partial \epsilon}
\end{array}\right]\right).
\end{aligned}
\end{equation}Substituting the value of $\boldsymbol{\mu}$ in \eqref{eq:FIM2}, after a few simple algebraic steps
\begin{equation}\label{eq:FIM_final}
\begin{aligned}
\boldsymbol{J}(\boldsymbol{\psi})=\frac{2}{\sigma^{2}}\left[\begin{array}{ccc}
L & \sum_{k=0}^{L-1} k &  \sum_{k=0}^{L-1} k^2  \\ \\%
 \sum_{k=0}^{L-1} k  &   \sum_{k=0}^{L-1} k^2  &  \sum_{k=0}^{L-1} k^3 \\\\%
  \sum_{k=0}^{L-1} k^2  &  \sum_{k=0}^{L-1} k^3  &  \sum_{k=0}^{L-1} k^4 
\end{array}\right].
\end{aligned}
\end{equation}Finally, the JCRB can be found by taking inverse of FIM, i.e., $\operatorname{JCRB}(\boldsymbol{\psi})=[\boldsymbol{J}(\boldsymbol{\psi})]^{-1}$ which is given in \eqref{eq:diagJBCR}. Hence, joint Cramer-Rao bounds of the unknown parameters \paramS are on the diagonal of the matrix $\operatorname{JCRB}(\boldsymbol{\psi})$

\begin{equation}\label{eq:diagJBCR}
\begin{aligned}
\operatorname{JCRB}(\boldsymbol{\psi})=[\boldsymbol{J}(\boldsymbol{\psi})]^{-1}=\frac{\sigma^{2}}{2} \left[
\begin{array}{ccc}
 \frac{9 (L-1) L+6}{L (L+1) (L+2)} & \frac{18-36 L}{L^3+3 L^2+2 L} & \frac{30}{L^3+3
   L^2+2 L} \\\\
 \frac{18-36 L}{L^3+3 L^2+2 L} & \frac{12 (2 L-1) (8 L-11)}{L^5-5 L^3+4 L} &
   -\frac{180}{L \left(L^3+L^2-4 L-4\right)} \\\\
 \frac{30}{L^3+3 L^2+2 L} & -\frac{180}{L \left(L^3+L^2-4 L-4\right)} &
   \frac{180}{L^5-5 L^3+4 L} \\
\end{array}
\right] .
\end{aligned}
\end{equation}\\

\subsection{Weighted Bayesian Cramer-Rao Bound (WBCRB)}

Weighted Bayesian CRBs take prior knowledge of parameters into consideration and are valid lower bounds for all signal-to-noise ratios. In \cite[Chapter~1]{van2007bayesian}, WBCRB is defined by using a weighted Fisher information matrix and a weighted Prior Information Matrix (PIM) as
\begin{equation}\label{eq:WBCRB}
\begin{aligned}
\operatorname{WBCRB}(\boldsymbol{\psi}) =\mathbb{E}_{\boldsymbol{\psi}}[\boldsymbol{Q}(\boldsymbol{\psi})]\left[\mathbb{E}_{\boldsymbol{\psi}}\left[\boldsymbol{J}_{d}(\boldsymbol{\psi})\right]+\mathbb{E}_{\boldsymbol{\psi}}\left[\boldsymbol{J}_{p}(\boldsymbol{\psi})\right]\right]^{-1} \mathbb{E}_{\boldsymbol{\psi}}[\boldsymbol{Q}(\boldsymbol{\psi})],
\end{aligned}
\end{equation}where $\boldsymbol{Q}(\boldsymbol{\psi})$ is a matrix for selected weighting functions. $\boldsymbol{J}_{p}(\boldsymbol{\psi})$ denotes the weighted PIM, which encompasses the prior distribution of the unknown parameters. $\boldsymbol{J}_{d}(\boldsymbol{\psi})$ denotes the weighted FIM. These terms are defined by following approach of \cite{van2007bayesian, NasirTime} as
\begin{equation}\label{eq:Q_func}
\begin{aligned}
\boldsymbol{Q}(\boldsymbol{\psi}) \overset{\Delta}{=} \operatorname{diag}\left(q(\theta), q(\omega), q(\epsilon)\right)
\end{aligned}
\end{equation}
\begin{equation}\label{eq:Jp_func}
\left[\boldsymbol{J}_{p}(\boldsymbol{\psi})\right]_{i j} \overset{\Delta}{=}  q\left(\psi_{i}\right) q\left(\psi_{j}\right) \frac{\partial \ln \left[p(\psi_i) q\left(\psi_{i}\right)\right]}{\partial \psi_{i}} \frac{\partial \ln \left[p(\psi_j) q\left(\psi_{j}\right)\right]}{\partial \psi_{j}}
\end{equation}

\begin{equation}\label{eq:Jd_func}
\left[\boldsymbol{J}_{d}(\boldsymbol{\psi})\right]_{i j} \overset{\Delta}{=}  q\left(\psi_{i}\right) q\left(\psi_{j}\right)[\boldsymbol{J}(\boldsymbol{\psi})]_{i j}.
\end{equation}As prior distribution of parameters \paramS is the uniform distribution, as described in Section~\ref{sec:system_mode}, $\theta \in (-\pi,\pi), \omega \in (-\omega_m,\omega_m)$ and $\epsilon \in (-\epsilon_m,\epsilon_m).$ We can write prior distributions as $p(\theta) = \frac{1}{2\pi}, p(\omega) = \frac{1}{2\omega_m}$ and $p(\epsilon) = \frac{1}{2\epsilon_m}.$\\
\subsubsection{$\boldsymbol{Q}(\boldsymbol{\psi}) $ and $\mathbb{E}_{\boldsymbol{\psi}}[\boldsymbol{Q}(\boldsymbol{\psi})]$}
By following \cite{van2007bayesian}, we select weighting functions as following
\begin{equation}\label{eq:Qi_func}
\begin{aligned}
q(\theta) &= \left(\frac{1}{2\pi}\right)^{2h} (\pi + \theta)^{h} (\pi - \theta)^{h} \\
q(\omega) &= \left(\frac{1}{2 \omega_m}\right)^{2h} (\omega_m + \omega)^{h} (\omega_m - \omega)^{h} \\
q(\epsilon) &= \left(\frac{1}{2 \epsilon_m}\right)^{2h} (\epsilon_m + \epsilon)^{h}(\epsilon_m - \epsilon)^{h} 
\end{aligned}
\end{equation}where $h$ is a weighting index which is used to control the tightness of the bound. After putting values from \eqref{eq:Qi_func} into \eqref{eq:Q_func} and by taking expectation, we can write
\begin{equation}\label{eq:E_Q_func}
\mathbb{E}_{\boldsymbol{\psi}}[\boldsymbol{Q}(\boldsymbol{\psi})] = 2^{-1-2h}\mathrm{B}(\frac{1}{2}, 1+h)\boldsymbol{I}_{3}
\end{equation}where $\mathrm{B}(x, y)=\int_{0}^{1} z^{x-1}(1-z)^{y-1} dz$ is the beta function.\\
\subsubsection{$\mathbb{E}_{\boldsymbol{\psi}}\left[\boldsymbol{J}_{p}(\boldsymbol{\psi})\right]$}
After putting values of prior distributions ($p(\theta), p(\omega)$ and $p(\epsilon)$) and weighting function from \eqref{eq:Qi_func} into \eqref{eq:Jp_func} and by taking expectation, we can write
\begin{equation}\label{eq:E_Jp_func}
\mathbb{E}_{\boldsymbol{\psi}}\left[\boldsymbol{J}_{p}(\boldsymbol{\psi})\right] = h\mathrm{B}(2h+1, 2h-1)\boldsymbol{P_3},
\end{equation}where
$$
\boldsymbol{P_3} = \operatorname{diag}\left(\left(\frac{1}{2\pi}\right)^{2}, \left(\frac{1}{2\omega_m}\right)^{2}, \left(\frac{1}{2\epsilon_m}\right)^{2}\right).
$$\\

\subsubsection{$\mathbb{E}_{\boldsymbol{\psi}}\left[\boldsymbol{J}_{d}(\boldsymbol{\psi})\right]$}
By using FIM defined in \eqref{eq:FIM_final} and weighting function from \eqref{eq:Qi_func} into \eqref{eq:Jd_func}, we can write
\begin{equation}\label{eq:E_Jd_func}
\mathbb{E}_{\boldsymbol{\psi}}\left[\boldsymbol{J}_{d}(\boldsymbol{\psi})\right] = \frac{2}{\sigma^{2}}\left[\begin{array}{ccc}
\lambda_{1}L & \lambda_{2}\sum_{k=0}^{L-1} k &  \lambda_{2}\sum_{k=0}^{L-1} k^2  \\ \\%
 \lambda_{2}\sum_{k=0}^{L-1} k  &   \lambda_{1}\sum_{k=0}^{L-1} k^2  &  \lambda_{2}\sum_{k=0}^{L-1} k^3 \\\\%
  \lambda_{2}\sum_{k=0}^{L-1} k^2  &  \lambda_{2}\sum_{k=0}^{L-1} k^3  &  \lambda_{1}\sum_{k=0}^{L-1} k^4 
\end{array}\right],
\end{equation}where 
$$
\lambda_{1} = \frac{h2^{-4h}}{0.5 +2h} \mathrm{B}(\frac{1}{2}, 2h)
$$ and 
$$
{\lambda_{2} = 4^{-1-2h} \left(\mathrm{B}(\frac{1}{2}, 1+h)\right)^{2}}.
$$Now, we can find WBCRB by using the closed-form expressions of $\mathbb{E}_{\boldsymbol{\psi}}[\boldsymbol{Q}(\boldsymbol{\psi})], \mathbb{E}_{\boldsymbol{\psi}}\left[\boldsymbol{J}_{p}(\boldsymbol{\psi})\right]$ and $\mathbb{E}_{\boldsymbol{\psi}}\left[\boldsymbol{J}_{d}(\boldsymbol{\psi})\right]$ in \eqref{eq:WBCRB}. \\

\section{Numerical Results}\label{sec:Numerical}
In this section, we verify the analysis by simulations of the proposed algorithm. We consider a communication system with Binary Phase Shift Keying (BPSK) modulation over AWGN channel. The instantaneous phase model is given in \eqref{eq:thetamodelQuadratic}. We used a regular low density parity code (LDPC) PEGReg252x504 of code rate 0.5 with $K = 252$ information bits and $N_c = 504$ coded bits \cite{mackay,Xiao-Yu2005}. After BPSK modulation we have 504 data symbols. With 30 preamble symbols, the total burst length is 534 symbols. We assumed that ${\theta_{-1}\sim \mathcal{U}(-\pi, \pi)}$, ${\omega_{-1}\sim \mathcal{U}(-0.01, 0.01)}$ and $\epsilon_{-1}\sim \mathcal{U}(-10^{-5}, 10^{-5})$ (unless stated otherwise). For fine-tuning of particles described in Section \ref{sec:FineTune}, we numerically compared the mean-square error (MSE) of the estimated parameters for different values of $\alpha$, $\zeta$ and $\gamma$ and we chose $\alpha = 0.1$, $\zeta = 0.01$ and $\gamma = 0.1 |\epsilon_{m}|$, as these values gave us minimum MSE of the estimated parameters. 

\subsection{Convergence of particles}

\begin{figure}
	\centering
  	\includegraphics[width=0.5\linewidth]{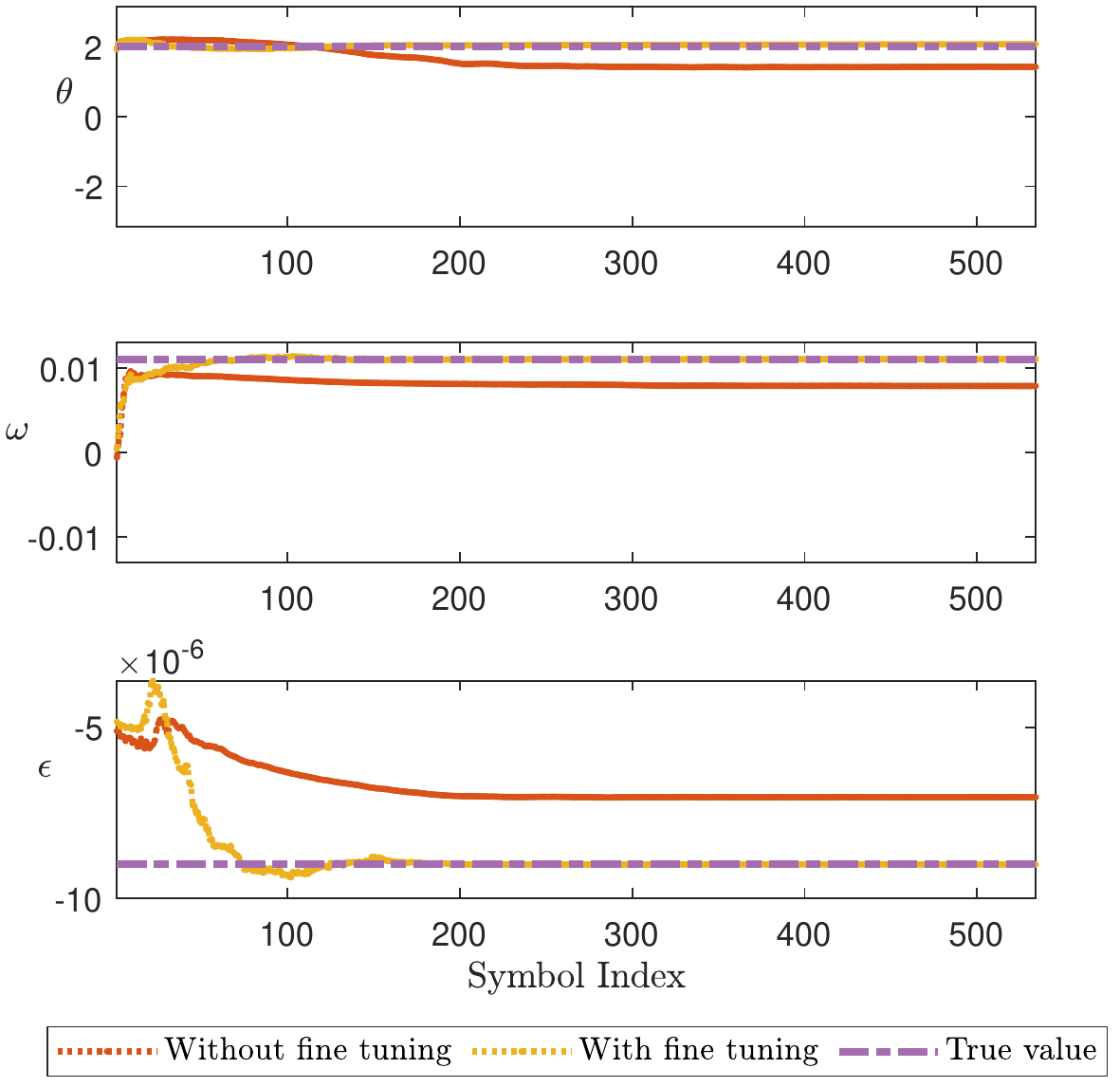}
  	\caption{Estimated $\hat{\theta}, \hat{\omega}$ and $\hat{\epsilon}$ at different time instants at $\text{SNR}=8$ dB for 400 particles (with and without fine-tuning of particles).}
\label{fig:param_estimate}
\end{figure}

In this section, we study the convergence of particles with and without fine-tuning for proposed particle filter algorithm. Figure~\ref{fig:param_estimate} presents estimates of $\theta$, $\omega$ and $\epsilon$, averaged over 100 bursts at different time instants when the SNR is set to $8$~dB and $400$ particles are used. The true values of parameters are $\theta = 2$, $\omega = 0.011$ and $\epsilon = \num{-9d-6}$. We assumed that ${\theta_{-1}\sim \mathcal{U}(-\pi, \pi)}$, ${\omega_{-1}\sim \mathcal{U}(-0.02, 0.02)}$ and $\epsilon_{-1}\sim \mathcal{U}(-10^{-5}, 0).$

It is evident that without fine-tuning, the estimate of $\omega$ converges to the wrong value after 30 symbols and in comparison particles of $\epsilon$ take a longer time to converge to the actual value. Hence, after a few instances, the particle filter for constant parameters fails to explore the sample space; the residual errors are the result of constant parameters lacking dynamics \cite{Kantas}. We are able to solve the problem of slow convergence of particles of $\epsilon$ by fine-tuning of particles and avoid the particles convergence to the wrong values of parameters. As shown in Figure~\ref{fig:param_estimate}, the particles of $\omega$ and $\epsilon$ converges to actual values at 150 symbols. Hence, fine-tuning converges the particles of $\omega$ and $\epsilon$ quickly and to the closest of the actual values of the parameters.

\subsection{Joint and Weighted Bayesian Cramer-Rao Bounds}

In this section, we compare the data-aided joint CRB and the weighted Bayesian CRB for ${L=534}$ under different prior distributions of $\theta$, $\omega$ and $\epsilon.$ For WBCRB, we used the weighting index $h = 1$ as it gives the tightest WBCRB \cite{van2007bayesian}. 

 \begin{figure}
	\centering
  	\includegraphics[width=0.95\linewidth]{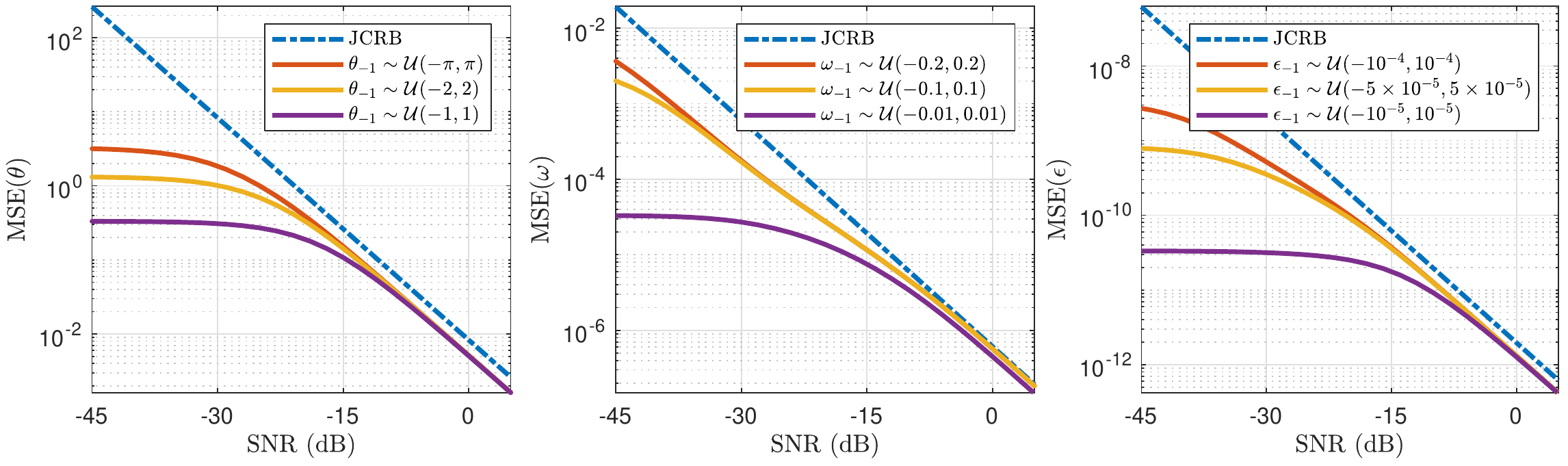}
  	\caption{JCRB and WBCRB for $L=534$.}
  	\label{fig:WBCRB_JCRB}
\end{figure}

The simulated results are depicted in Fig.~\ref{fig:WBCRB_JCRB}, where JCRB is shown in dotted line and WBCRBs with different prior distributions of parameters are shown in solid lines. At higher SNR, WCRB approaches JCRB. The SNR at which the WBCRB approaches the JCRB depends on the parameter estimation range. For the wider prior distribution of any parameter, the WBCRB approaches to the JCRB at very low SNR. For example, for ${\epsilon_{-1}\sim \mathcal{U}(-10^{-4}, 10^{-4})}$, WBCRB approaches JCRB at $-16$ dB. However, for narrower prior distribution ${\epsilon_{-1}\sim \mathcal{U}(-10^{-5}, 10^{-5})}$, WBCRB approaches JCRB at $-4$ dB SNR.

The WBCRBs of $\theta$, $\omega$ and $\epsilon$ goes flat at lower SNR. Each WBCRB converges to the variance of the prior distribution of respective parameter at lower SNR. For example, for ${\omega_{-1}\sim \mathcal{U}(-0.01, 0.01)}$, WBCRB converges to $\frac{1}{12}\left(0.02\right)^2 = \num{3.3d-5}$ and for ${\theta_{-1}\sim \mathcal{U}(-\pi, \pi)}$, WBCRB converges to $\frac{1}{12}\left(2\pi\right)^2 = \num{3.28}.$ \\

\subsection{Mean-Square Errors and Cramer-Rao Bounds}

 \begin{figure}
	\centering
  	\includegraphics[width=0.95\linewidth]{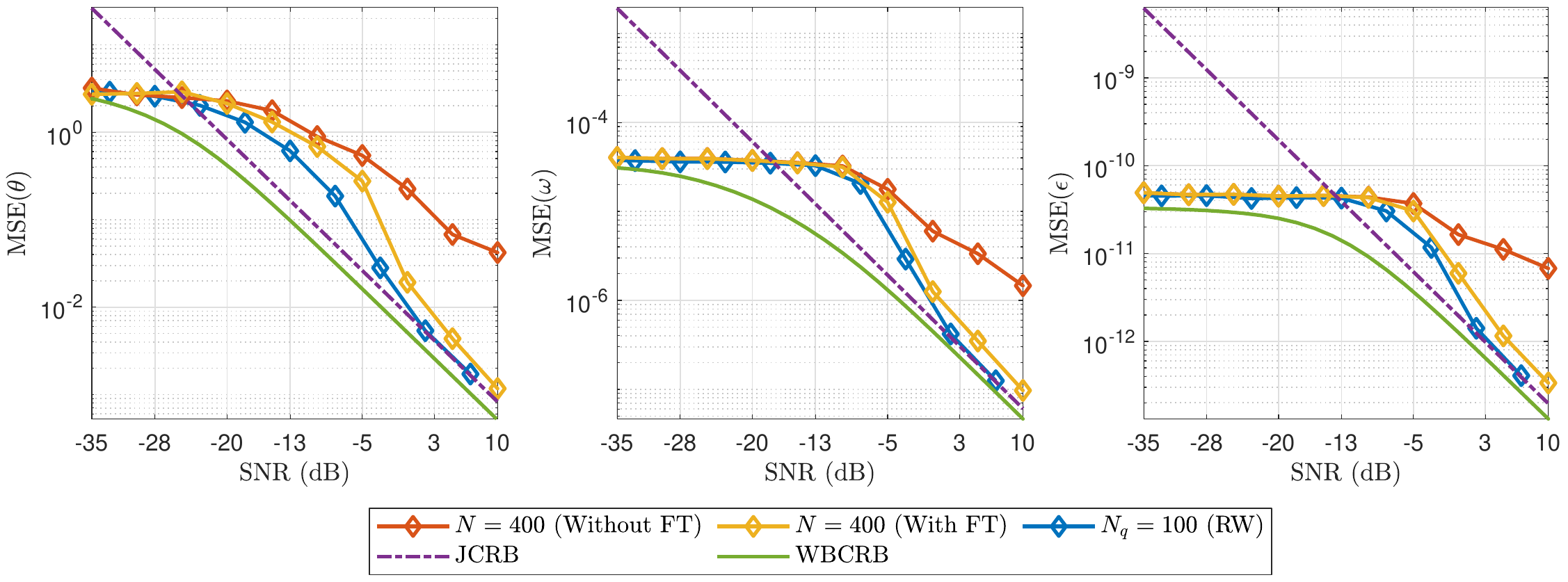}
  	\caption{Comparison of semi data-aided MSE and data-aided JCRB and WBCRB for $L=534$.}
  	\label{fig:WBCRB_JCRB_MSE}
\end{figure}

In this section, we simulate the mean-square error of semi data-aided estimations of $\theta$, $\omega$ and $\epsilon$ using particle filter (with and without fine-tuning) and random-walk phase model based approximation. We run the simulations with same parameters as described at the start of the Section \ref{sec:Numerical}. In these simulations we further consider only one receive node and no iteration is performed between decoder and estimator. Phase tracking by random walk model followed by unwrapping and curve fitting is performed with $N_q = 100$ phase quantization levels. For particle filtering, we used $N=400$ particles. To plot the JCRB and WBCRB, we assumed that all the data symbols are known. 

From Fig.~\ref{fig:WBCRB_JCRB_MSE}, it is evident that at lower SNRs, MSE of estimated parameters is lower than the JCRB and approaches to the WBCRB. Thus, WBCRB is a valid bound for all SNRs, however, JCRB is tighter at higher SNR. It depicts that WBCRB is a valid lower bound for all SNR values, however, the WBCRB is not as tight as JCRB at high SNRs.

Further, we have two main observations from Fig.~\ref{fig:WBCRB_JCRB_MSE}; first, the fine-tuning of particles provides a significant performance improvement. For example, we achieve 10 dB SNR gain by fine-tuning of particles for $\epsilon$ at MSE of $\num{5.9d-12}$, compared to when no fine-tuning is performed. Secondly, the proposed algorithm which uses random-walk phase model based approximation have better performance than particle filter method. The MSE of all parameters for RW (with $N_q =100$) approaches the JCRB at 2 dB SNR. We achieve nearly 2 dB SNR gain for RW based algorithm for $\omega$ at MSE of $\num{1d-7}$, compared to the particle filter method with fine-tuning.

\subsection{Bit Error Rate (BER) performance}
In this section we evaluate the BER performance with proposed algorithms and study the impact of different parameters on BER performance. \\

\subsubsection{PF with and without fine-tuning} 
\begin{figure}
	\centering
  	\includegraphics[width=0.48\linewidth]{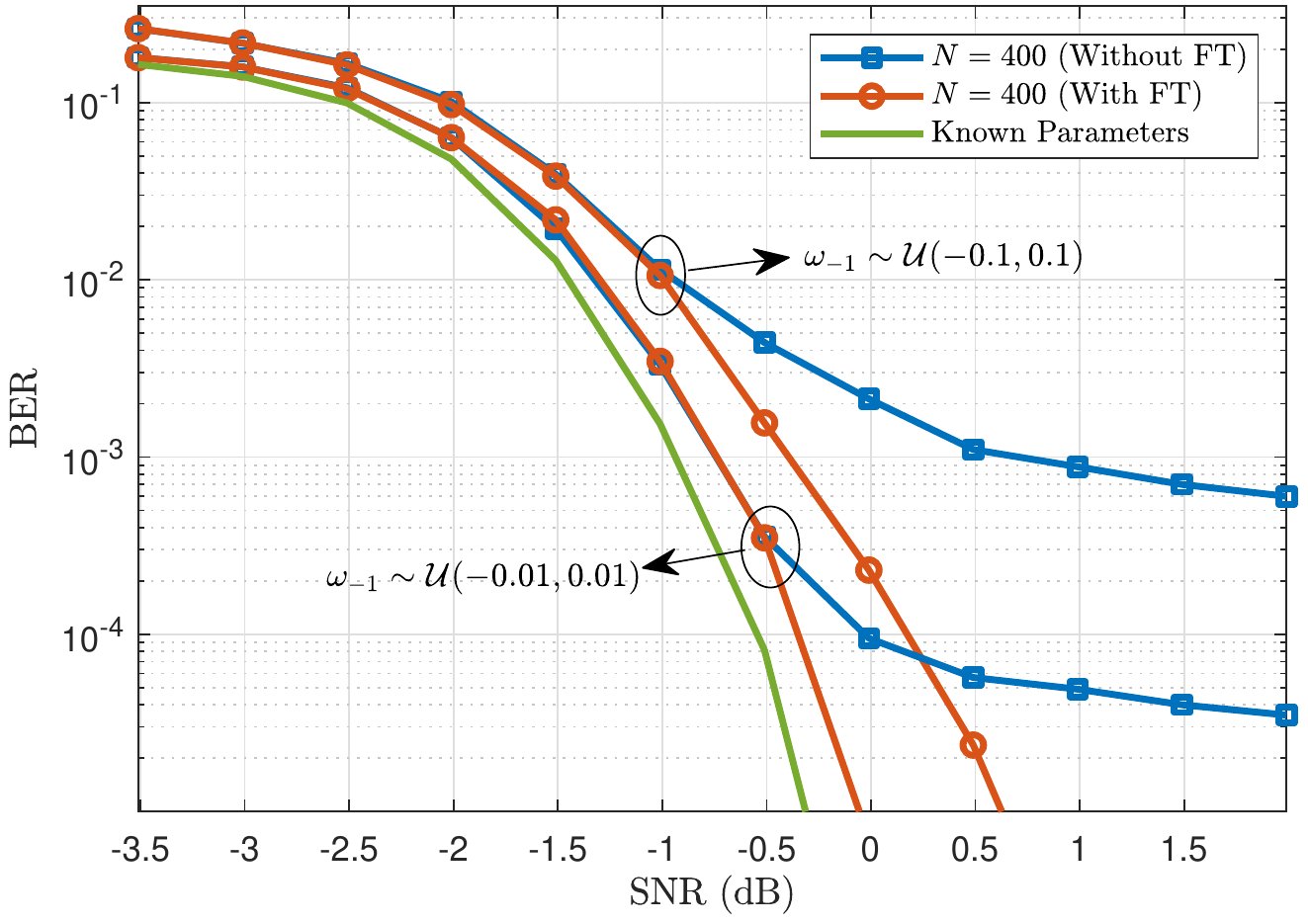}
  	\caption{Comparison of BER performance with and without fine-tuning for the particle filter method.}
  	\label{fig:ber_FT_NT}
\end{figure}

In Fig.~\ref{fig:ber_FT_NT}, we compare the particle filter method with and without fine-tuning of particles at two prior distributions {of $\omega$}, i.e., ${\omega_{-1}\sim \mathcal{U}(-0.01, 0.01)}$ and ${\omega_{-1}\sim \mathcal{U}(-0.1, 0.1)}.$ We used 400 particles. In these simulations we further consider only one receive node and no iteration is performed between decoder and estimator.

It is evident from Fig.~\ref{fig:ber_FT_NT} that the fine-tuning of particles improves the BER performance. Without fine-tuning of particles, at higher SNRs, BER does not decrease quickly. As we have shown in Fig.~\ref{fig:param_estimate}, the reason is that without fine-tuning, the particles of $\theta$ and $\omega$ converge quickly to a value and particle of $\epsilon$ take a longer time to converge which results in residual errors in estimation $\omega$ and $\epsilon$, which are very sensitive synchronization parameters, causing larger bit errors. At BER $\num{7d-4}$ when ${\omega_{-1}\sim \mathcal{U}(-0.1, 0.1)}$, we achieve nearly $2.3$ dB SNR gain. Similarly, At $\num{3d-5}$ BER when ${\omega_{-1}\sim \mathcal{U}(-0.01, 0.01)}$, we achieve nearly $2.25$ dB SNR gain.

One more observation we can make is that the gap between ideal performance and particle filter method's BER performance with fine-tuning increases as the prior distribution of $\omega$ becomes more wider. For example, fine-tuning achieves $\num{2d-5}$ BER at nearly $-0.1$ dB when ${\omega_{-1}\sim \mathcal{U}(-0.01, 0.01)}$ and the same BER for ${\omega_{-1}\sim \mathcal{U}(-0.01, 0.01)}$ is achieved at SNR of $0.5$ dB.\\

\subsubsection{Impact of number of particles} Next, we discuss the impact of number of particles on BER performance of particle filter. For these simulations, we assumed ${\omega_{-1}\sim \mathcal{U}(-0.1, 0.1)}$ and SNR is $0$ dB. In these simulations we further consider only one receive node and no iteration is performed between decoder and estimator.

\begin{figure}
	\centering
  	\includegraphics[width=0.48\linewidth]{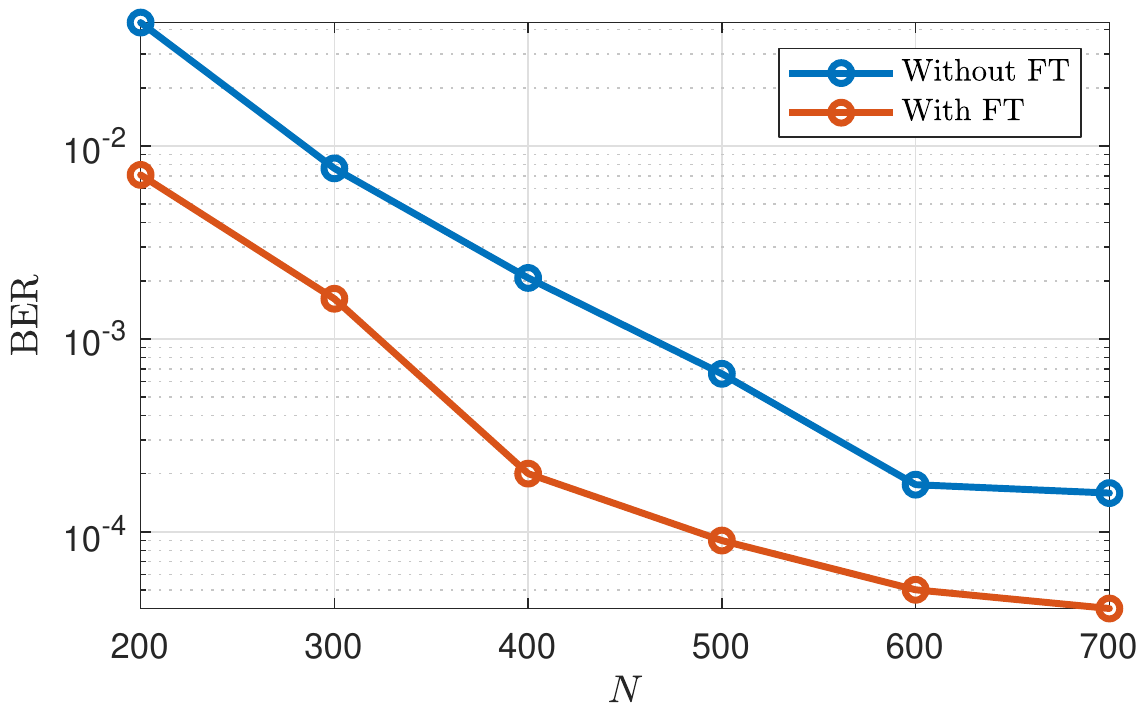}
  	\caption{Effect of number of particles on BER performance at SNR of {$0$ dB}.}
  	\label{fig:ber_NoPF}
\end{figure}

The BER performance of particle filtering with and without fine-tuning is plotted for with respect to the number of particles in Fig.~\ref{fig:ber_NoPF}. It is evident that as number of particles increase BER performance improves. With or without fine-tuning, we achieve nearly 1 decade improvement in BER performance when the number of particles is increased from 300 to 500. We can achieve same BER performance for 400 particles with fine-tuning as of 600 particles without fine-tuning. For example,  we achieve $\num{2d-4}$ BER at 400 particles with fine-tuning and 600 particles without fine-tuning. Note that, without fine-tuning of particles we do not get any significant performance improvement after $N=600$ because higher number of particles do not improve convergence speed of particles of $\epsilon$. \\

\subsubsection{Impact of number of global iterations} 
\begin{figure}
	\centering
  	\includegraphics[width=0.48\linewidth]{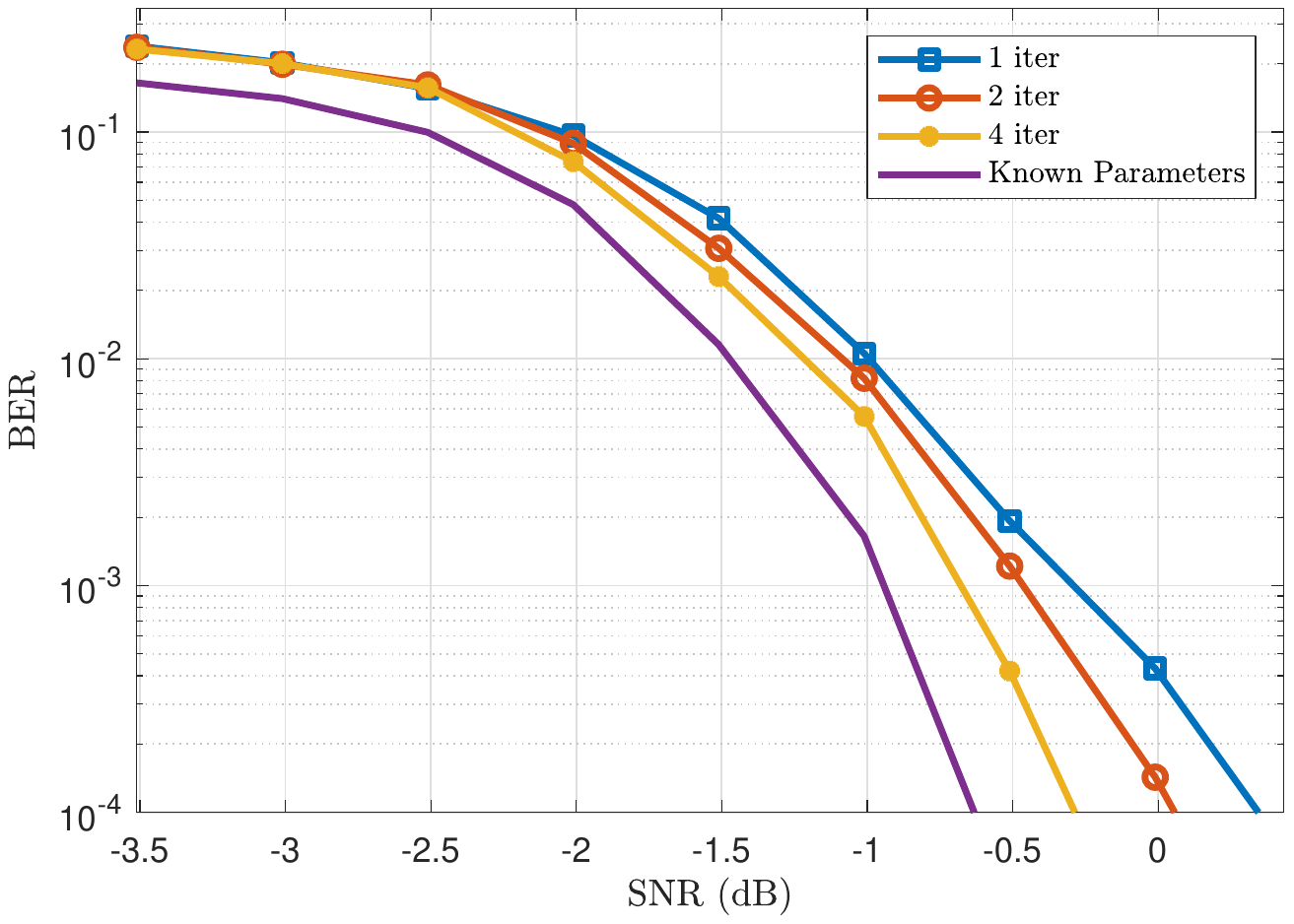}
  	\caption{Effect of number of global iterations on BER performance at different SNRs for RW phase model based technique with ${N_q=100}.$}
  	\label{fig:ber_iter}
\end{figure}

Next, we discuss the impact of number of global iterations between estimator and decoder on BER performance of one node. For these simulations, we assumed ${\theta_{-1}\sim \mathcal{U}(-\pi, \pi)}$, ${\omega_{-1}\sim \mathcal{U}(-0.03, 0.03)}$ and $\epsilon_{-1}\sim \mathcal{U}(-10^{-6}, 10^{-6})$. We performed the phase tracking by random walk model followed by unwrapping and curve fitting by using $N_q = 100$ phase quantization levels. The BER performance is shown in Fig.~\ref{fig:ber_iter}. It is evident that as number of global iterations increases BER performance improves. For example,  we achieve 0.5 dB SNR gain with 4 iterations at $\num{4d-4}$ BER.\\

\subsubsection{PF vs RW for multiple receive nodes} 
\begin{figure}
	\centering
  	\includegraphics[width=0.48\linewidth]{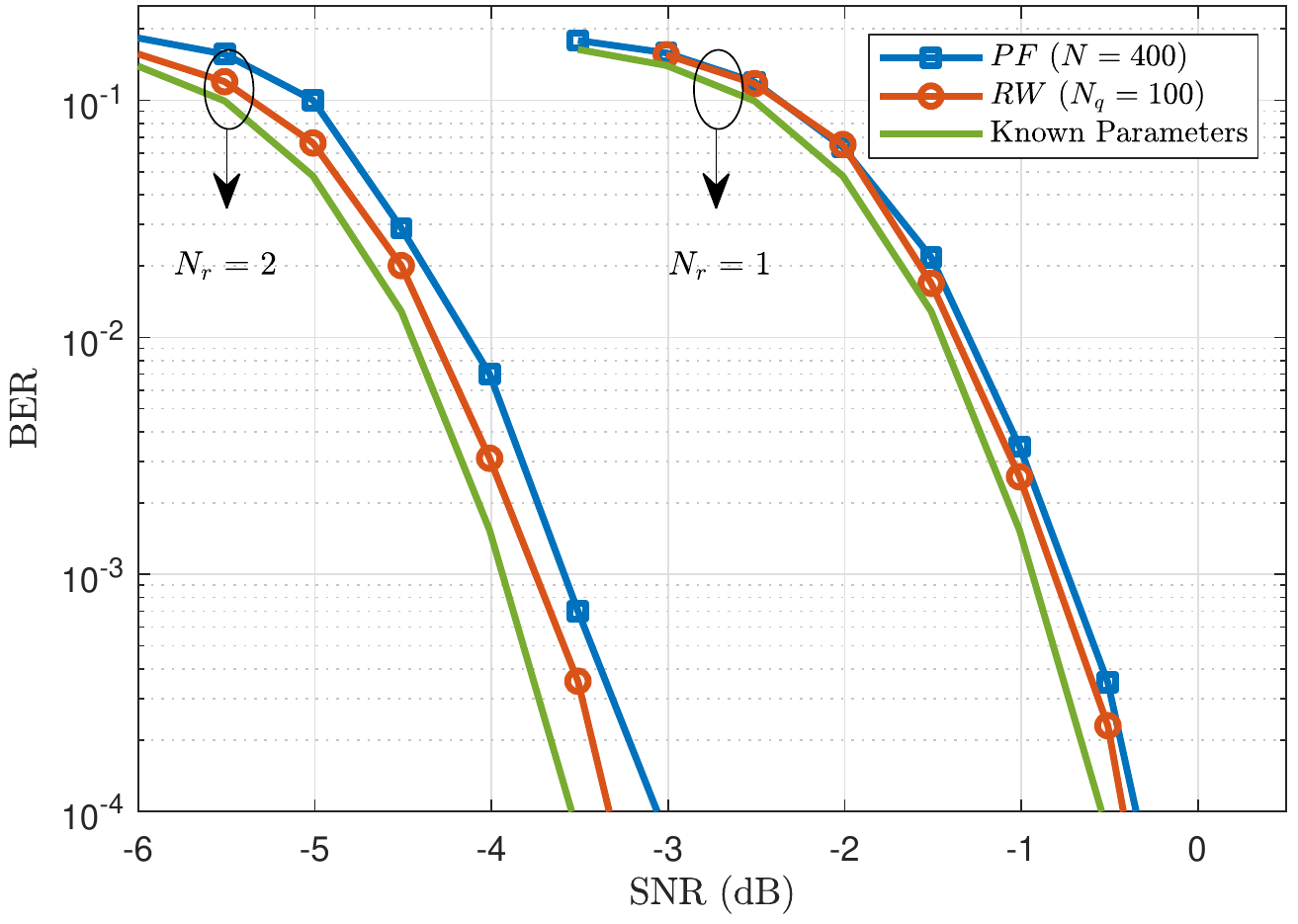}
  	\caption{Comparison of BER performance for proposed algorithms for two nodes.}
  	\label{fig:ber_performance}
\end{figure}

In Fig.~\ref{fig:ber_performance}, we compare the particle filter and random-walk based model for multiple receive nodes. We used number of particles $N = 400$ for the particle filter with fine-tuning. Phase tracking by random walk model followed by unwrapping and curve fitting is performed with $N_q = 100$ phase quantization levels. We performed only one global iteration between estimation part and decoding.


The gap between PF and ideal BER performance is slightly larger in comparison to random walk phase model. When we increase the number of receive nodes then at $\num{2d-3}$ BER, we achieve nearly $3$ dB gain in SNR for random walk phase model and for particle filter we achieve $2.7$ dB SNR gain. Note that, a gain of $3$ dB is the maximum achievable gain using two receiver nodes. 

\section{Conclusion}\label{sec:conclusion}
In this work for a distributed receiver, we presented an algorithm for estimating joint channel parameters (carrier phase, Doppler shift, and Doppler rate) and decoding iteratively decodable codes transmitted over channels affected by Doppler shift and Doppler rate. Sum-product algorithms are applied to factor graphs to derive this algorithm. We have proposed two methods for dealing with intractable integrals. For the estimation of unknown parameters, we employed particle filtering with sequential importance sampling (SIS). With the second approach, a random walk phase model was used to approximate our model followed by a phase tracking algorithm and polynomial regression algorithm to estimate the unknown parameters. We also proposed a method for fine-tuning particles for particle filtering methods to improve their convergence speed. We derived the Weighted Bayesian Cramer-Rao Bounds (WBCRB) for joint carrier phase, Doppler shift and Doppler rate estimation, which takes into account the prior distribution of the estimation parameters and is an accurate lower bound for all considered Signal to Noise Ratio values. The Monte Carlo simulations are run to study the proposed algorithm's bit error rate (BER) performance.


\appendix

\section{Derivation of (14)}\label{Appendix1}
By employing Bayes theorem and laws of conditional probability, we obtain a recursive decomposition of \eqref{eq:posterior_pdf_factor} as following.
\begin{equation}\label{eq:posterior_pdf_factor1}
\begin{aligned}
p\left(\paramInEqSI \mid y_{0: k}\right) 
&=\frac{p\left(\paramInEqSI, y_{0: k}\right)}{p\left(y_{0: k}\right)} =\frac{p\left(\paramInEqSI, y_{k}, y_{0: k-1}\right)}{p\left(y_{k}, y_{0: k-1}\right)} \\
&=\frac{p\left(\paramInEqSI, y_{k} \mid y_{0: k-1}\right) p\left(y_{0: k-1}\right)}{p\left(y_{k} \mid y_{0: k-1}\right) p\left(y_{0: k-1}\right)} \\
&=D_{k} \times p\left( \paramInEqSI, y_{k} \mid y_{0: k-1}\right) \\
&=D_{k} \times p\left(y_{k} \mid  \paramInEqSI, y_{0: k-1}\right) \times p\left(\paramInEqSI \mid y_{0: k-1}\right)
\end{aligned}
\end{equation}where $D_{k} = p\left(y_{k}\mid y_{0: k-1}\right)^{-1}.$ Given the current state of parameters, the current observation only relies on the current state and is independent from the previous observations, so, we have $ {p\left(y_{k} \mid \paramInEqSI \right)=p\left(y_{k} \mid \theta_{k}^{(i)}, \omega_{k}^{(i)}, \epsilon_{k}^{(i)}\right)}$. So, we can write target distribution as

\begin{equation}\label{eq:posterior_pdf_factor2}
\begin{aligned}
p\left(x_{0: k}^{(i)}, a_{0: k}^{(i)}, b_{0: k}^{(i)}, c_{0: k}^{(i)} \mid y_{0: k}\right) &=D_{k} \times p\left(y_{k} \mid \theta_{k}^{(i)}, \omega_{k}^{(i)}, \epsilon_{k}^{(i)}\right) \times p\left(\theta_{k}^{(i)}, \omega_{k}^{(i)}, \epsilon_{k}^{(i)},  \paramInEqSIK \mid y_{0: k-1}\right)\\
&=D_{k} \times p\left(y_{k} \mid \theta_{k}^{(i)}, \omega_{k}^{(i)}, \epsilon_{k}^{(i)}\right) \times  p\left(\theta_{k}^{(i)}, \omega_{k}^{(i)}, \epsilon_{k}^{(i)} \mid  \paramInEqSIK, y_{0: k-1}\right) \\
& \times p\left(\paramInEqSIK \mid y_{0: k-1}\right) \\
\end{aligned}
\end{equation}As $\theta_{k}^{(i)}, \omega_{k}^{(i)}$ and $\epsilon_{k}^{(i)}$ are independent, so we can write

\begin{equation}\label{eq:posterior_pdf_factor3}
\begin{aligned}
p\left(\theta_{k}^{(i)}, \omega_{k}^{(i)}, \epsilon_{k}^{(i)} \mid  \paramInEqSIK, y_{0: k-1}\right) &=  p\left(\theta_{k}^{(i)} \mid  \paramInEqSIK, y_{0: k-1}\right) \times  p\left(\omega_{k}^{(i)} \mid  \paramInEqSIK, y_{0: k-1}\right) \\
&\times p\left( \epsilon_{k}^{(i)} \mid  \paramInEqSIK, y_{0: k-1}\right)
\end{aligned}
\end{equation}By using \eqref{eq:posterior_pdf_factor3} in \eqref{eq:posterior_pdf_factor2}, we get same expression as given in \eqref{eq:posterior_pdf_factor}. This concludes our derivation.


\section*{Acknowledgment}
This work has been supported by the SmartSat CRC, whose activities are funded by the Australian Government’s CRC Program. This work has been supported by an Australian Government Research Training Program (RTP) Stipend and RTP Fee-Offset Scholarship through University of South Australia.

\ifCLASSOPTIONcaptionsoff
  \newpage
\fi

\bibliographystyle{IEEEtran}
\bibliography{arXiv_Joint_iterative_EstDecoding_SPA.bib}
%
%
%
%
%

\end{document}